\documentclass[aps,prx,superscriptaddress,twocolumn,longbibliography, notitlepage]{revtex4-1}

\usepackage{graphicx}
\usepackage{amsmath}
\usepackage{amssymb}
\usepackage{amsfonts}
\usepackage{amsthm}
\usepackage{hyperref}
\usepackage{color}
\usepackage{soul}
\usepackage{textcomp}
\usepackage{bm}
\usepackage{dsfont}
\usepackage{relsize}
\usepackage{braket}
\usepackage{enumitem}   
\usepackage{mathrsfs}
\usepackage{cleveref}

\theoremstyle{plain}

\numberwithin{obs}{section}

\newcommand{\comments}[1]{}
\newcommand{\ba}{\begin{align}}
\newcommand{\ea}{\end{align}}

\newcommand{\blue}[1]{\textcolor{black}{#1}}

\newcommand{\ee}{{\mathcal{E}}}

\newcommand{\TrS}[1]{\text{Tr}_{\scriptscriptstyle S}\big(#1\big)}
\newcommand{\TrB}[1]{\text{Tr}_{\scriptscriptstyle B}\big(#1\big)}

\newcommand{\sys}[1]{#1_{\scriptscriptstyle S}}
\newcommand{\bath}[1]{#1_{\scriptscriptstyle B}}

\newcommand{\Tr}[1]{\text{Tr}\left(#1\right)}
\newcommand{\TrSB}[1]{\text{Tr}_{\scriptscriptstyle SB}\left(#1\right)}
\newcommand{\SB}[1]{#1_{\scriptscriptstyle SB}}
\newcommand{\SBD}[1]{#1_{\scriptscriptstyle SBD}}
\newcommand{\De}[1]{#1_{\scriptscriptstyle D}}
\newcommand{\average}[2]{\left \langle #1 \right \rangle_{#2}}
\newcommand{\rom}[1]{\uppercase\expandafter{\romannumeral #1\relax}}

\newcommand{\sqrbra}[1]{\left[ #1\right]}
\newcommand{\curbra}[1]{\left\{ #1\right\}}
\newcommand{\lind}{\mathscr{L}}

\newcommand{\timeorderedexp}{\overleftarrow{\mathcal{T}}{\text{exp}}}

 \newcommand{\Real}{\text{Re}}
  \newcommand{\operatorS}{\mathbb{S}}
  \newcommand{\Wdis}{W_{\rm diss}}

 \def\be{\begin{eqnarray}}
\def\ee{\end{eqnarray}}
\renewcommand{\d}{\mbox{d}}
\usepackage{mathptmx,framed}
\usepackage{bigints}
\usepackage[normalem]{ulem}

\newcommand{\tr}{\mbox{tr}}
\renewcommand{\d}{\mbox{d}}
\def\be{\begin{eqnarray}}
\def\ee{\end{eqnarray}}

\newcommand{\diff}{\mathcal{I}_t(\pi_t,\dot{H}_t)} % was: \mathcal{I}_t(\pi_t,\dot{H}_t)
\newcommand{\diffA}{\mathcal{I}_t(\pi_t,A)}
\newcommand{\diffX}{\mathcal{I}_t(\pi_t, X)} % was: \mathcal{I}_t(\pi_t, X)

\begin{document}

\title{Work fluctuations in slow processes: quantum signatures and optimal control}
\author{Harry J. D. Miller}
\thanks{These authors contributed equally to this work.}
\affiliation{Department of Physics and Astronomy, University of Exeter, Stocker Road, Exeter EX4 4QL, UK.}
\author{Matteo Scandi}
\thanks{These authors contributed equally to this work.}
\affiliation{Max-Planck-Institut f\"ur Quantenoptik, D-85748 Garching, Germany}\affiliation{ICFO-Institut de Ciencies Fotoniques, The Barcelona Institute of\\
	Science and Technology, Castelldefels (Barcelona), 08860, Spain}
\author{Janet Anders}
\affiliation{Department of Physics and Astronomy, University of Exeter, Stocker Road, Exeter EX4 4QL, UK.}
\author{Mart\'i Perarnau-Llobet}
\affiliation{Max-Planck-Institut f\"ur Quantenoptik, D-85748 Garching, Germany}
\date{\today}

\begin{abstract}
An important result in classical stochastic thermodynamics is the work fluctuation--dissipation relation (FDR), which states that the dissipated work done along a slow process is proportional to the resulting work fluctuations. 
Here we show that slowly driven quantum systems violate this FDR whenever quantum coherence is generated along the protocol, and derive a quantum generalisation of the work FDR. The additional quantum terms in the FDR are found to lead to a non-Gaussian work distribution. Fundamentally, our result shows that quantum fluctuations prohibit finding slow protocols that minimise both dissipation and fluctuations simultaneously, in contrast to classical slow processes. Instead, we develop a quantum geometric framework to find processes with an optimal trade-off between the two quantities.
\end{abstract}

%The additional quantum terms of the  FDR are shown to imply a non-Gaussian work distribution. 

\maketitle

Thermodynamics traditionally deals with macroscopic systems at thermal equilibrium, and its laws relate averages of quantities such as work and heat. When bringing the theory to the microscale, fluctuations become
significant and can no longer be neglected with respect to average quantities.  As a consequence, a stochastic description of thermodynamic processes is needed, which has triggered enormous attention to the understanding of  work  (and heat) fluctuations
~\cite{Esposito2009,Jarzynski2011,Seifert_2012,Campisi2011c}. In the regime of slow but finite-time classical processes, work fluctuations are governed by a single relation, known as the work fluctuation-dissipation-relation (FDR)~\cite{Hermans1991,Wood1991,Hendrix,Jarzynski1997d}:
\begin{align}\label{eq:FDR}
\Wdis=\frac{1}{2}\beta\sigma_{w}^2.
\end{align}
Here,  $\sigma_w^2 \equiv \langle w^2 \rangle - \langle w \rangle^2 $ is the variance of the work distribution $P(w)$ and $\Wdis \equiv \langle w \rangle -\Delta F \geq 0$ the average dissipated work \blue{along the protocol}, i.e. the difference between average work done and the change of equilibrium free energy $\Delta F$, which is always non-negative due to the second law, \blue{and $\beta=1/k_B T$ with $T$ the inverse temperature of the environment.}
The work FDR \eqref{eq:FDR}  is one of the pillars of classical stochastic thermodynamics; it shows that near equilibrium work fluctuations are responsible for dissipation, and conversely that any optimal slow process that minimises dissipation will subsequently minimise the fluctuations~\cite{Crooks2007,Sivak2012a}.
\blue{For many slow classical processes the work distribution $P(w)$ is Gaussian~\cite{Speck,Chvosta2007,_ubrt_2007,Speck_2011,Hoppenau_2013}, and if the process also fulfils Jarzynski's equality then this immediately implies Eq.~\eqref{eq:FDR}~\cite{Jarzynski1997d}. }

\blue{For quantum systems, developing a definition of work and understanding how quantum effects influence its statistics has raised much attention recently~\cite{Talkner2007c,Talkner2009,Jarzynski2015,Talkner2016,
Miller2017,Perarnau2017,Solinas2017,Aberg2018,baumer2018fluctuating,
Holmes2019,Wu2019}. Previous studies on the non-classicality of \blue{work distributions} have considered the  emergence of quasiprobabilities due to weak measurement~\cite{Allahverdyan2014,Solinas2015,PatrickPotts2019}, contextuality~\cite{Lostaglio2018}, and violations of macrorealism \cite{Blattmann2017,Miller2018a}. Despite the wealth of research on this topic, a  quantum generalisation of~\eqref{eq:FDR} has not  been addressed.} 

\blue{\blue{Based initially on the Two-Projective-Measurement (TPM)  distribution $P(w)$ ~\cite{Talkner2007c,Talkner2009,Esposito2009},} in this article we derive a quantum work FDR and find that it differs from~\eqref{eq:FDR} through an additional contribution arising due to  quantum fluctuations generated along the protocol. This extra term is positive-definite implying that  slow quantum processes are governed by the \emph{inequality} $\Wdis\leq \beta\sigma_{w}^2/2$, with equality obtained when no coherences in energy are created during the dynamics. We further demonstrate that the extra quantum term in the FDR leads to a non-Gaussian $P(w)$, and show that the same quantum FDR is also valid for work distributions accessed from weak measurements of the system.}

While quantum work fluctuations are of fundamental interest, understanding their behaviour also provides a method for minimising them in practical implementations. Indeed, the design of reliable and minimally-dissipative thermodynamic engines  is of utmost importance in quantum thermodynamics. In the  regime of slow processes, the minimisation of dissipation can be obtained using techniques from differential geometry: one can equip the thermodynamic state space with a Riemannian metric~\cite{Weinhold1975,Ruppeiner1979}, and optimal protocols can be found by calculating the associated geodesics~\cite{Nulton1,Salamon1,Crooks2007,Sivak2012a,Zulkowski2013,Zulkowski2015,Deffner2018,Scandi}. Here, we show  that the quantum work fluctuations can also be related to a Riemannian metric. However, due to quantum modifications this new metric only coincides with the metric responsible for minimising dissipation in the classical commutative regime. While this result  rules out  protocols that simultaneously minimise both $\Wdis$ and $\sigma_w$ for quantum coherent processes, our framework can be used to find optimal trade-offs between dissipation and fluctuations. 

These results are derived under three main assumptions: 
(i)~the coupling between  system and bath is weak, 
%, which leads to the standard Born-Markov approximation.
(ii)~the system reaches thermal equilibrium when interacting with the bath,
(iii)~the driving is slow, so that we can  expand the magnitudes of interest in the driving velocity and keep only leading terms. 
%in the sense that we can expand the magnitudes of interest in $v$ and keep only  orders of $1/\tau$, where $\tau$ is the total time of the process, and keep only leading order terms (here $\tau$ should be compared with the time of scale of equilibration). 
Under these assumptions, we now  derive a quantum version of the FDR in Eq.~\eqref{eq:FDR}.

\textit{The quantum work FDR.} We study the thermodynamics of an open quantum system S coupled to a thermal bath B  with total Hamiltonian $\SB{H}(t)=\sys{H}(t)+\bath{H}+\SB{V}$, where $\sys{H}(t)=\sys{H}(t)\otimes\bath{\mathbb{I}}$ is the driven system Hamiltonian and $\SB{V}$ a small but finite coupling Hamiltonian. We take a finite-time interval  $t\in [0,\tau]$ and consider processes where  the two system Hamiltonian endpoints are fixed, $\sys{H}(0)=H_0$ and $\sys{H}(\tau)=H_\tau$. 
We assume that the initial density matrix of S and B is a product $\SB{\varrho}(0)=\sys{\pi}(0)\otimes \bath{\pi}(0)$ where $\sys{\pi}(0)=e^{-\beta \sys{H}(0)}/\sys{Z}(0)$ and $\bath{\pi}=e^{-\beta \bath{H}}/\bath{Z}$ are the respective Gibbs states for the bare system and bath.  The compound system evolves as $\SB{\varrho}(t)=U(t) \, \SB{\varrho}(0) \, U^{\dagger}(t)$ with the time-ordered exponential $U(t)=\timeorderedexp\big(-(i/\hbar)\int^t_0 dt' \ \SB{H}(t')\big)$. 
\blue{Work is required to perform $U(t)$, and because only the system Hamiltonian changes in time while coupling is weak, this work can be associated with work on the system alone \cite{Talkner2009}. The work  statistics can be defined via the TPM scheme, where  a projective energy measurement of the total Hamiltonian  is performed at the beginning, $\SB{H}(0)$, and the end, $\SB{H}(\tau)$, of the process, with the energy differences measured identified as the work values $w$. From the statistics, the work distribution can then be constructed and becomes $P(w) = {1 \over 2\pi} \int d\lambda \ e^{-i\lambda w} \, G(\lambda)$ with a moment generating function $G(\lambda) =   \TrSB{U^{\dagger}({\tau}) e^{i\lambda \SB{H}({\tau})} U({\tau}) e^{-i\lambda \SB{H}({0})} \SB{\varrho}(0)}$~\cite{Talkner2007c,Talkner2009,Esposito2009}, which directly gives the work moments via $\langle w^k \rangle=(-i)^{k} (d^k/d\lambda^k)G(\lambda)\big|_{\lambda=0}$. While in the following we will use the TPM work distribution to establish the quantum FDR, we show in Appendix C that our results are also valid for alternative work distributions based on weak measurements~\cite{Allahverdyan2014,Solinas2015,PatrickPotts2019}.
}

From now on we shall use the more compact notation  $X_t \equiv \sys{X}(t)$, with $X=\varrho, H ,\pi$ and denote $\Tr{.}$ as the trace over the system degrees of freedom. 
\textcolor{black}{In general, the reduced dynamics of the system can be written as $\dot{\varrho}_t = - {i \over \hbar} \, \TrB{[\SB{H}(t),\SB{\varrho}(t)]}=\mathscr{L}_t[\varrho_t]$. Here, we will assume that the system follows an adiabatic Markovian master equation with a unique instantaneous steady state given by the thermal state at each  $ t\in[0,\tau]$: $\mathscr{L}_t[\pi_t]=0,$ with $\pi_t=e^{-\beta H_t}/Z_t$ (a precise form of $\mathscr{L}_t[\varrho_t]$ is presented in Appendix D). This is well--justified whenever the bath dynamics are fast compared to the driving rate of the system Hamiltonian~\cite{Albash2012,Dann},  and the coupling between S and B is weak enough to satisfy the Born-Markov approximation and the rotating wave approximation~\cite{breuer2002theory}.  %These assumptions apply to a wide range of weakly-coupled open quantum systems.
 Importantly, under these assumptions the TPM statistics can be determined by unravelling the master equation in terms of quantum jump trajectories~\cite{Silaev2014a,Liu2016a,Liu2016b}. These trajectories can then be accessed via local measurements of a quantum detector~\cite{Pekola2013b}, circumventing the need to perform global energy measurements.}
Under these assumptions, we show in Appendix A that the work fluctuations $\sigma^2_w \equiv \langle w^2 \rangle - \langle w \rangle^2$ are given by
\begin{align}\label{eq:fluctuations}
\sigma^2_w=2\int^\tau_0\hspace{-2mm} dt_{1} \int^{t_{1}}_0 \hspace{-2mm}dt_{2} \ \Tr{\dot{H}_{t_{1}}\overleftarrow{P}(t_{1},t_{2})\left[\operatorS_{\varrho_{t_{2}}}(\dot{H}_{t_{2}})\right]},
\end{align}
where $\overleftarrow{P}(t_1,t_2)=\overleftarrow{\mathcal{T}}\text{exp}\left(\int_{t_2}^{t_1}d\nu \ \mathscr{L}_\nu \right)$ is the propagator for the Lindbladian, and we have introduced the linear mapping 
\begin{align}\label{eq:Jop}
\operatorS_\varrho(A):=\frac{1}{2}\lbrace\varrho, \Delta_\varrho A\rbrace, \ \ \ 
\end{align}
with $\Delta_\varrho A=A-\Tr{A\varrho}$ \blue{and $\{ , \}$ denoting the anticommutator}. 
We now assume that the total time $\tau$ of the process is large with respect to the time scale(s) of thermalisation, which are encoded in $\mathscr{L}_t$. Since the two endpoints of the trajectory are fixed at $H_0$ and $H_\tau$, one has $\dot{H}_t\propto \tau^{-1}$. In this case, we can expand the relevant expressions in terms of $\tau^{-1}$ and keep the leading orders, which we refer to as the slow driving regime. This assumption allows us to further simplify Eq.~\eqref{eq:fluctuations} in Appendix B, using techniques similar to the ones developed in  \cite{Mandal2016a} for classical systems. To first order in $\tau^{-1}$ the work fluctuations are
\begin{align}
\label{eq:fluctuationsQS}
\sigma^2_w \simeq  -2\int_{0}^{\tau}dt \,\Tr{\dot H_{t} \, \lind^{+}_{t} \sqrbra{\operatorS_{\pi_t}(\dot{H}_{t})}} .
\end{align}
Note that the integrand is proportional to $\tau^{-2}$, and so for the whole integral $\sigma^2_w  \propto \tau^{-1}$ as desired. In Eq.~\eqref{eq:fluctuationsQS},  we have introduced the so-called Drazin inverse $\mathscr{L}^+_t$ of the Lindblad operator $\mathscr{L}_t$ \cite{Boullion1971,Scandi}. This inverse is defined as 
\begin{align}\label{eq:Drazin}
\mathscr{L}^+_t[A]:=\int^\infty_0 d\nu \ e^{\nu\mathscr{L}_t}\big[\pi_t \, \Tr{A}-A\big],
\end{align}
and satisfies three conditions \cite{Scandi}: 
\quad (i) commutation with the Lindbladian, i.e. $\mathscr{L}_t\mathscr{L}^+_t[A]=\mathscr{L}^+_t\mathscr{L}_t[A]=A-\pi_t\Tr{A}$, \quad (ii) invariance of the thermal state, i.e. $\mathscr{L}^+_t[\pi_t]=0$, 
\quad and (iii) tracelessness, i.e. $\Tr{\mathscr{L}^+_t[A]}=0$.

An expression similar to Eq.~\eqref{eq:fluctuationsQS}  describes the dissipated work, $\Wdis$, in slow quantum processes~\cite{Cavina2017,Scandi} 
\begin{align}\label{eq:diss}
\Wdis = - \beta \, \int_{0}^{\tau}dt \, \Tr{\dot H_{t} \, \lind^{+}_{t} \sqrbra{\mathbb{J}_{\pi_{t}}(\dot{H}_{t})}}.
\end{align}
Note, that in place of  $\operatorS_{\pi_t}$ in Eq.~\eqref{eq:fluctuationsQS} the map $\mathbb{J}_{\pi_{t}}$ appears, with  
\begin{align}\label{eq:JopII}
\mathbb{J}_\varrho(A):=\int^1_0 da \,\, \, \varrho^a \,\, \, \Delta_\varrho A \,\, \, \varrho^{1-a}.
\end{align}
In the special case that $A$ commutes with $\varrho$ the maps $\operatorS_\varrho(A)$ and $\mathbb{J}_\varrho(A)$ both reduce to $\operatorS_\varrho(A) = \varrho \, \, \Delta_\varrho A = \mathbb{J}_\varrho(A)$.

\textcolor{black}{Taking the expressions for work fluctuations, $\sigma^2_w $, and the dissipated work, $\Wdis$, together, we obtain the quantum work FDR:
\begin{align}\label{eq:qFDR}
\frac{1}{2}\beta\sigma^2_w =\Wdis + Q_w ,
\end{align}
where $Q_w=\beta\int^\tau_0 dt \ \Tr{\dot{H}_t \,\, \lind^{+}_{t}\sqrbra{(\mathbb{J}_{\pi_{t}}-\operatorS_{\pi_{t}})(\dot{H}_t)}}$  is a quantum correction coming from the difference between the maps $\operatorS_\varrho(A)$ and $\mathbb{J}_\varrho(A)$. 
}

\blue{In  Appendix D we show that $Q_w \geq 0$, with equality  if and only if  $[\dot{H}_t,H_t]=0$  for  $\beta>0$ and $\forall t\in[0,\tau]$.  This implies that for  slow quantum processes with $[H_t,\dot{H}_t] \not=0$ the classical FDR  in Eq.~\eqref{eq:FDR} breaks down and the work fluctuations are in fact greater than dissipation. In general, one has an \textit{inequality}:
\begin{align}\label{eq:FDRqq}
	\Wdis\leq \frac{1}{2} \, \beta \, \sigma_{w}^{2} .
\end{align}
We can then interpret the quantum work FDR \eqref{eq:qFDR}  as follows:  during a slow process where the state remains close to a thermal state $\pi_t$,  the work fluctuations $\beta \sigma^2_{w}/2$ can be divided into two positive contributions: a thermal contribution $\Wdis$, which arises from the thermal fluctuations in $\pi_t$, and a purely quantum contribution $Q_w$, which appears whenever quantum fluctuations are created in the dynamics as  $[\pi_t, \dot{H}_t]\neq 0$.}

Let us rewrite $Q_w=\beta\int^\tau_0 dt \diff$ where we have introduced the \textit{dynamical skew information} \newline $\diffA := \Tr{A \,\, \lind^{+}_{t}\sqrbra{(\mathbb{J}_{\pi_{t}}-\operatorS_{\pi_{t}})(A)}}$ for an arbitrary observable $A$. To further elaborate the idea that  $Q_w$ measures the  quantum work fluctuations, for now suppose S evolves under a perfectly thermalising map with a single time-scale $1/\Gamma$, i.e. the Lindbladian satisfies
\begin{align}
\mathscr{L}_t[\varrho_t]= (\pi_t-\varrho_t) \, \Gamma,
\label{eq:Lt}
\end{align}
which has the Drazin inverse $\mathscr{L}_t^+(.)=(\Tr{.}\pi_t- \mathbb{I}(.))/\Gamma$. In this case, $\diffA$  becomes proportional to the average Wigner-Yanase-Dyson skew information~\cite{Hansen2008,Frerot2016,Miller2018}: $\diffA =-\frac{1}{2\Gamma}\int^1_0 da \, \, \Tr{[A,\pi^a] \,[A,\pi^{1-a}]}$ which can be understood as a measure of quantum uncertainty in the observable $A$ \cite{Luo2005}: it is positive and vanishes \textit{iff} $[A,\pi_t]=0$, reduces to the usual variance for pure $\pi_t=\ket{\psi}\bra{\psi}$, and decreases under classical mixing. For more general Lindbladians, $\diffA$  also takes into account  the presence of different timescales of thermalisation  through the additional dependence on $\mathscr{L}_t^+$.  Summarising, in Eq.~\eqref{eq:qFDR} we can interpret $Q_w$ as a measure of the time-integrated quantum fluctuations in the power $\dot{H}_t$.

\textit{Non-Gaussianity of the work distribution.} Here we show that these quantum coherences necessarily lead to a non-Gaussian shape of the TPM work distribution $P(w)$. For this $P(w)$ the Jarzynski equality holds \cite{Talkner2009}, which relates the change in equilibrium free energy to the cumulants of work done on the system that are computed from $P(w)$:
\begin{align}\label{eq:jarz}
\Delta F=-\beta^{-1}\ln \langle e^{-\beta w} \rangle=\sum^\infty_{k=1}\frac{(-\beta)^{k-1}}{k!}\kappa^{(k)}_{ w}.
\end{align}
Here $\kappa^{(k)}_{ w}$ are the cumulants of work, with $\kappa^{(1)}_{ w}=\langle w \rangle$ and $\kappa^{(2)}_{ w}=\sigma^2_w$. After rearranging terms in~\eqref{eq:jarz} and combining this with the quantum FDR~\eqref{eq:qFDR}, we find
\begin{align}\label{eq:cumulants}
\sum^\infty_{k=3}\frac{(-\beta)^{k-1}}{k!}\kappa^{(k)}_{ w}=Q_w\geq 0.
\end{align}
In fact, as we have seen, $Q_w$ vanishes \textit{iff} $[\dot{H}_t,H_t]=0$ for \newline $\forall t\in[0,\tau]$. Since a Gaussian work distribution has zero cumulants for $k\geq3$, we conclude that $P(w)$ necessarily becomes non-Gaussian whenever the process generates coherences of the power operator with respect to the instantaneous Hamiltonian. 
\blue{This contrasts with the classical expectation that slow processes lead to Gaussian work distributions \cite{Hendrix,Speck}.} 
\blue{The equality~\eqref{eq:cumulants} further demonstrates that measuring the work cumulants can provide a direct witness of quantum fluctuations in power. }

\medskip

\textit{Thermodynamic geometry and optimal paths.}
\blue{Now that we have established a relationship between work dissipation and fluctuations, we are in a position to determine optimal protocols. In order to find protocols with minimal fluctuations, one can take a geometric approach similar to~\cite{Scandi, Crooks2007, Sivak2012a}.} 

Considering a decomposition of the system Hamiltonian of the form $H_t=X_0+\vec{\lambda}_t \cdot \vec{X}$, where  $\vec{\lambda}_t=(\lambda_1(t), \lambda_2(t), ...)$ is the vector of scalar controllable parameters and \newline $\vec{X}=\partial H_t/\partial \vec{\lambda}_t=(X_1, X_2, ...)$ are the corresponding generalised conjugate forces. Then, Eq.~\eqref{eq:fluctuationsQS} can be recast in the form \newline $\sigma_w^{2}= {2 \over \beta} \int_{0}^{\tau}dt\,\ \left[\frac{d\vec{\lambda}_t}{dt}\right]^T  \bm{\Lambda}(\vec{\lambda}_t)   \left[\frac{d\vec{\lambda}_t}{dt}\right]$, 
where $\bm{\Lambda}(\vec{\lambda}_t)$ has elements 
\begin{align}\label{eq:metricFluct}
 \hspace{-2mm}	\Lambda_{ij}(\vec{\lambda}_t):=-\frac{\beta}{2} \, \Tr{X_i \, \, \mathscr{L}_t^+\left[\operatorS_{\pi_t}(X_j) \right] + X_j \, \, \mathscr{L}_t^+\left[\operatorS_{\pi_t}(X_i) \right]}.
\end{align}
It follows that since the rate of dissipated work and dynamical skew information are both positive, $\bm{\Lambda}(\vec{\lambda}_t)$ is a positive-definite matrix. Since $\bm{\Lambda}(\vec{\lambda}_t)$ is also symmetric and depends smoothly on $\pi_t$, it induces a Riemannian metric on the space of quantum thermal states~\cite{Petz1994a}.  
Differential geometry then provides an efficient and systematic approach to find optimal protocols by solving Euler-Lagrange equations for the functional $\sigma_w^{2}$ of the curve $\vec{\lambda}_t$. Curves of minimal fluctuations are identified as geodesics of constant velocity.

The work--fluctuation metric $\bm{\Lambda}(\vec{\lambda}_t)$ given in Eq.~\eqref{eq:metricFluct} should be compared to the work--dissipation metric $\bm{\xi}(\vec{\lambda}_t)$, for which $\Wdis= \int_{0}^{\tau}dt\,\ \big[\frac{d\vec{\lambda}_t}{dt}\big]^T  \bm{\xi}(\vec{\lambda}_t)   \big[\frac{d\vec{\lambda}_t}{dt}\big]$, with elements~\cite{Scandi}
\begin{align}\label{eq:metricDiss}
\hspace{-1mm}	\xi_{ij}(\vec{\lambda}_t):=-\frac{\beta}{2} \, \Tr{X_i \, \, \mathscr{L}_t^+\left[\mathbb{J}_{\pi_t}(X_j) \right] + X_j \, \, \mathscr{L}_t^+\left[\mathbb{J}_{\pi_t}(X_i) \right]}.  
\end{align}
The two metrics $\bm{\Lambda}(\vec{\lambda}_t)$ and $\bm{\xi}(\vec{\lambda}_t)$ coincide whenever the conjugate forces commute ie. $[X_i,X_0]=[X_i,X_j]=0 \ \forall i,j$. In this special case both metrics reduce to the classical Fisher-Rao metric over the space of thermal states, multiplied with $k_B T$ and an integral relaxation time related to the open system dynamics \cite{Sivak2012a}.

In general, the fluctuation and dissipation metrics differ and hence their corresponding geodesics will no longer coincide, in contrast to slow processes in classical thermodynamics. In other words,  for quantum processes, any slow protocol $\vec{\lambda}_t^{opt}$ that minimises dissipation will have non-minimal fluctuations, and vice versa. To interpolate between these two extremes, one can resort to minimising the objective function
\begin{align}\label{eq:alphaObjective}
	\mathcal{C}_\alpha:=\alpha \, \tilde{\sigma}_w^{2}+(1-\alpha) \, \Wdis \quad \mbox{for} \quad \alpha\in[0,1],
\end{align}
where $\alpha$ weights the relative importance of the fluctuations versus dissipation and  $\tilde{\sigma}^{2}_w=\frac{1}{2}\beta\sigma^2_w$.  The family of metrics minimising $\mathcal{C}_\alpha$ for weights $\alpha$ is just the convex sum $\bm{g}_\alpha(\vec{\lambda}_t) = \alpha \, \bm{\Lambda}(\vec{\lambda}_t) + (1-\alpha) \,  \bm{\xi}(\vec{\lambda}_t)$. In Appendix E we use Euler-Lagrange methods to find the optimal protocol $\lambda_t^{\text{opt}}(\alpha)$ that minimises $\mathcal{C}_\alpha$ when $\lambda_t$ is a one-dimensional control parameter with $H_t=X_0+\lambda_t X$. The optimal velocity takes the form $\dot{\lambda}_t^{\text{opt}}(\alpha)\propto \sqrt{\xi (\lambda_t) + \alpha \, \, \beta \diffX}$ which clearly depends on $\alpha$ due to the presence of quantum coherence. This contrasts with the classical case $[X_0,X]=0$ where the optimal protocol can be obtained for any $\alpha$ by driving the system at a constant dissipation rate \cite{Sivak2012a}.

\medskip

\begin{figure}[t]
	\centering
	\hspace{-0.3cm} \includegraphics[height=0.34\linewidth]{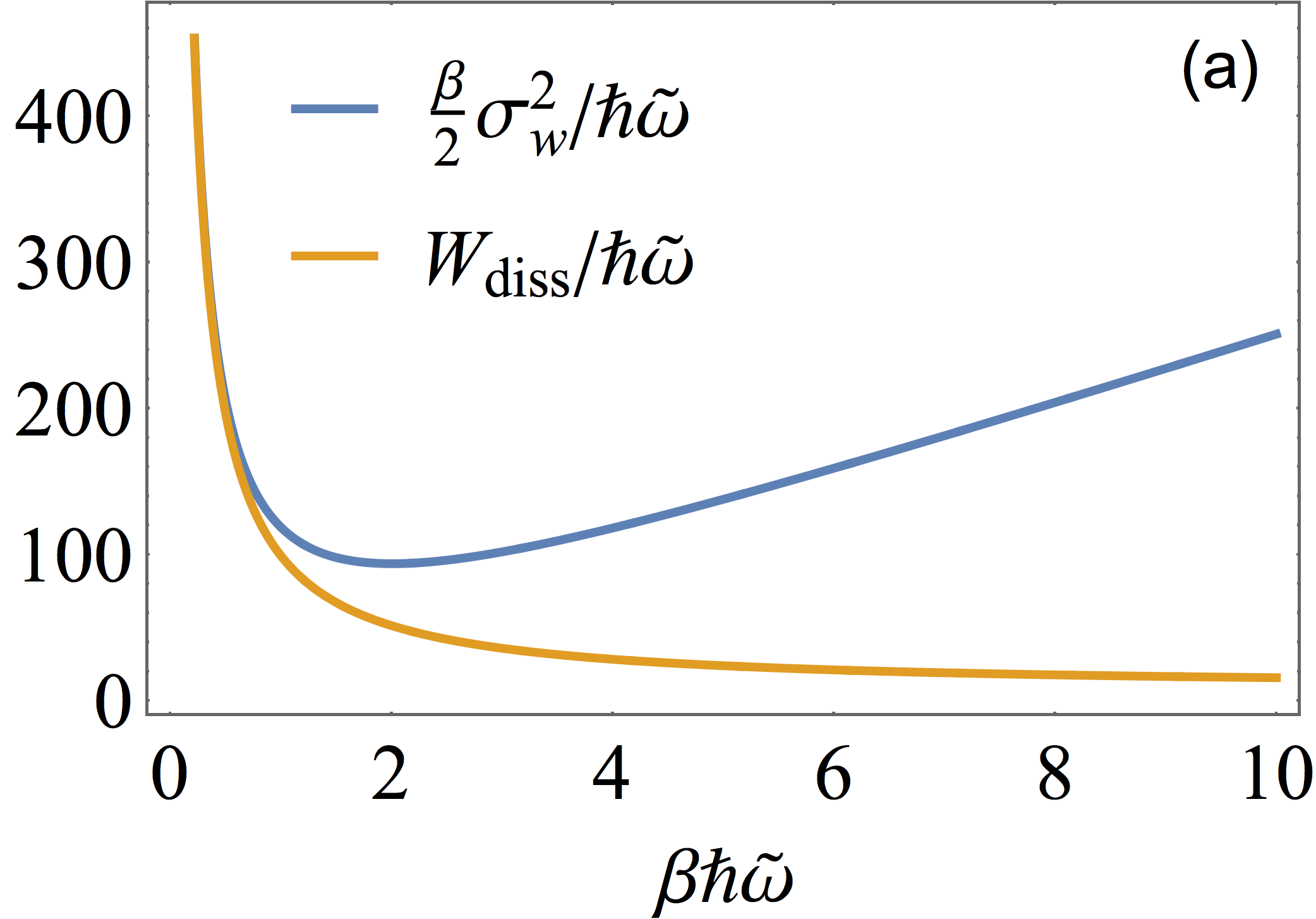}\, \quad
	\includegraphics[height=0.34\linewidth]{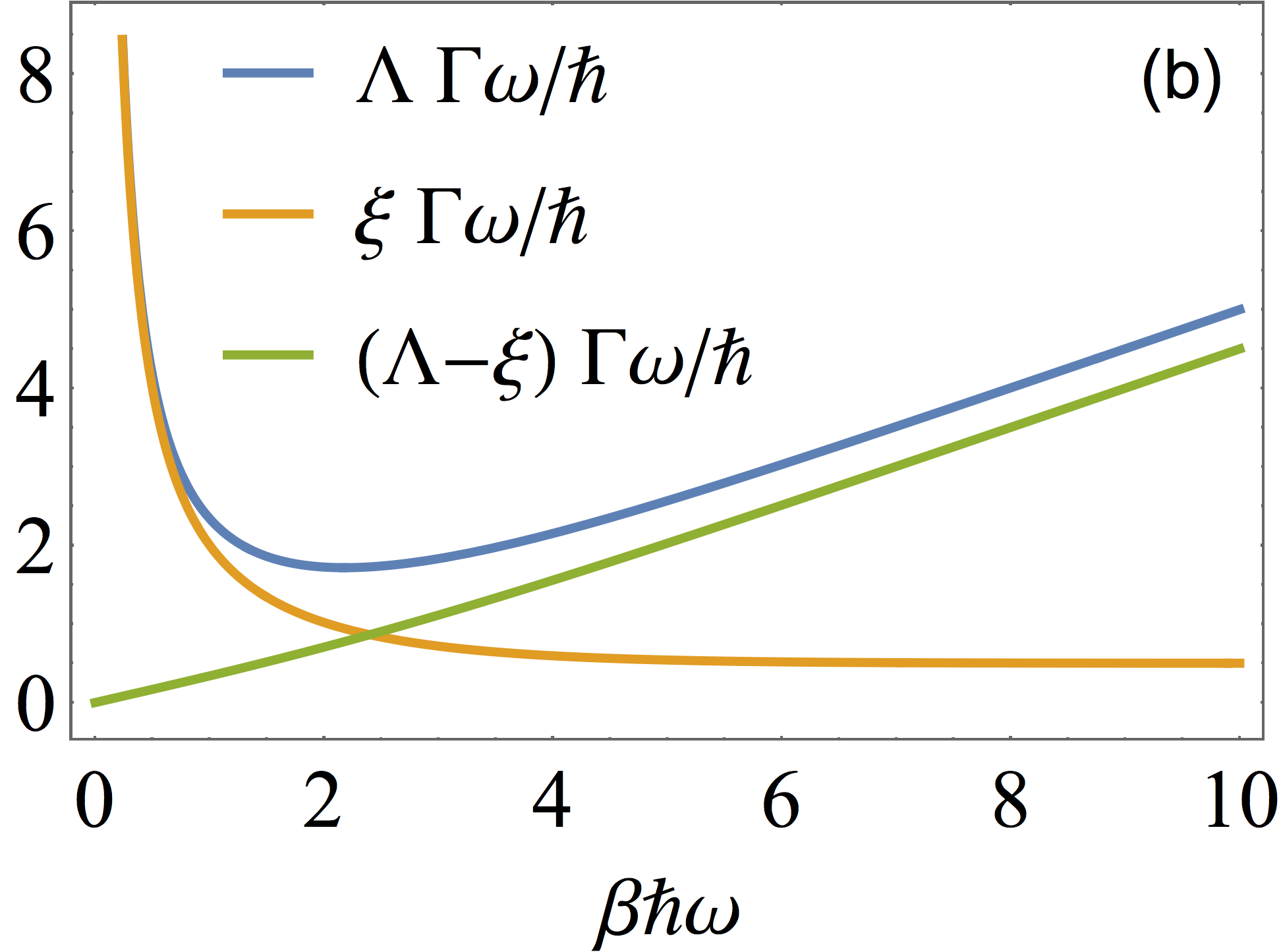}
	\caption{{\bf (a)} Dissipated work, $\Wdis$, and work fluctuations, $\frac{1}{2} \, \beta \, \sigma_{w}^{2}$, as a function of initial state inverse temperature $\beta$ for the harmonic oscillator example. The plots are for a protocol in which the oscillator frequency $\omega_t$ is increased linearly in time from $\omega_0 = 0.1 \tilde{\omega}$ to $\omega_1 = 10 \tilde{\omega}$ for a fixed reference frequency $\tilde{\omega}$.
{\bf (b)} Plot of the metric tensors of fluctuations ($\Lambda$),  dissipation ($\xi$) and of their difference ($\Lambda-\xi$),  for the harmonic oscillator example as a function of inverse temperature $\beta$ at a given energy gap $\hbar{\omega}$ (see Appendix F). 
}
	\label{fig:geod4c}
\end{figure}

\textit{Example.} Let us  illustrate our results with a slowly driven harmonic oscillator, $H_t= \hbar \omega_t \left(a^\dagger_{\omega_t}a_{\omega_t}+1/2\right)$, connected to a perfectly-thermalising bath described by the master equation Eq.~\eqref{eq:Lt}.  Here $\omega_t$ is the time-dependent frequency of the oscillator, and $a_{\omega_t}$ and $a^\dagger_{\omega_t}$ are the frequency-dependent creation and annihilation operators. Taking the time-derivative yields the  power operator $\dot{H}_t= \hbar\dot{\omega}_t (H_t/\hbar \omega_t+((a^\dagger_{\omega_t})^2+a^2_{\omega_t})/2)$,
which does not commute with the instantaneous Hamiltonian $H_t$, i.e. $[H_t,\dot{H}_t]\neq0$. In Fig.~\ref{fig:geod4c}(a), we compare the expressions for $\Wdis$ and  $\beta\sigma_{w}^{2}/2$ \textcolor{black}{for a slow linear ramp of $\omega_t$}, 
and it can be seen that the curves differ substantially at low temperatures (i.e. high $\beta$), where quantum fluctuations become dominant, and become closer for higher temperatures, where thermal fluctuations dominate and classical behaviour is recovered.
\blue{The corresponding metrics $\Lambda(\omega_t)$ and $\xi(\omega_t)$ along with their difference,  $\Lambda(\omega_t) - \xi(\omega_t) = \beta \diffX$, are shown in Fig.~\ref{fig:geod4c}(b) as a function of inverse temperature. As expected, this difference vanishes in the high temperature limit ($\beta \to 0$). In the low temperature regime thermal fluctuations given by the dissipation metric $\xi(\omega_t)$ decay, while quantum coherences contribute more significantly to the total fluctuations in power that are given by $\Lambda(\omega_t)$. The details of all these calculations are provided in Appendix F.}

\begin{figure}[t]
	\centering
\hspace{-0.5cm}	\includegraphics[width=1.05\linewidth]{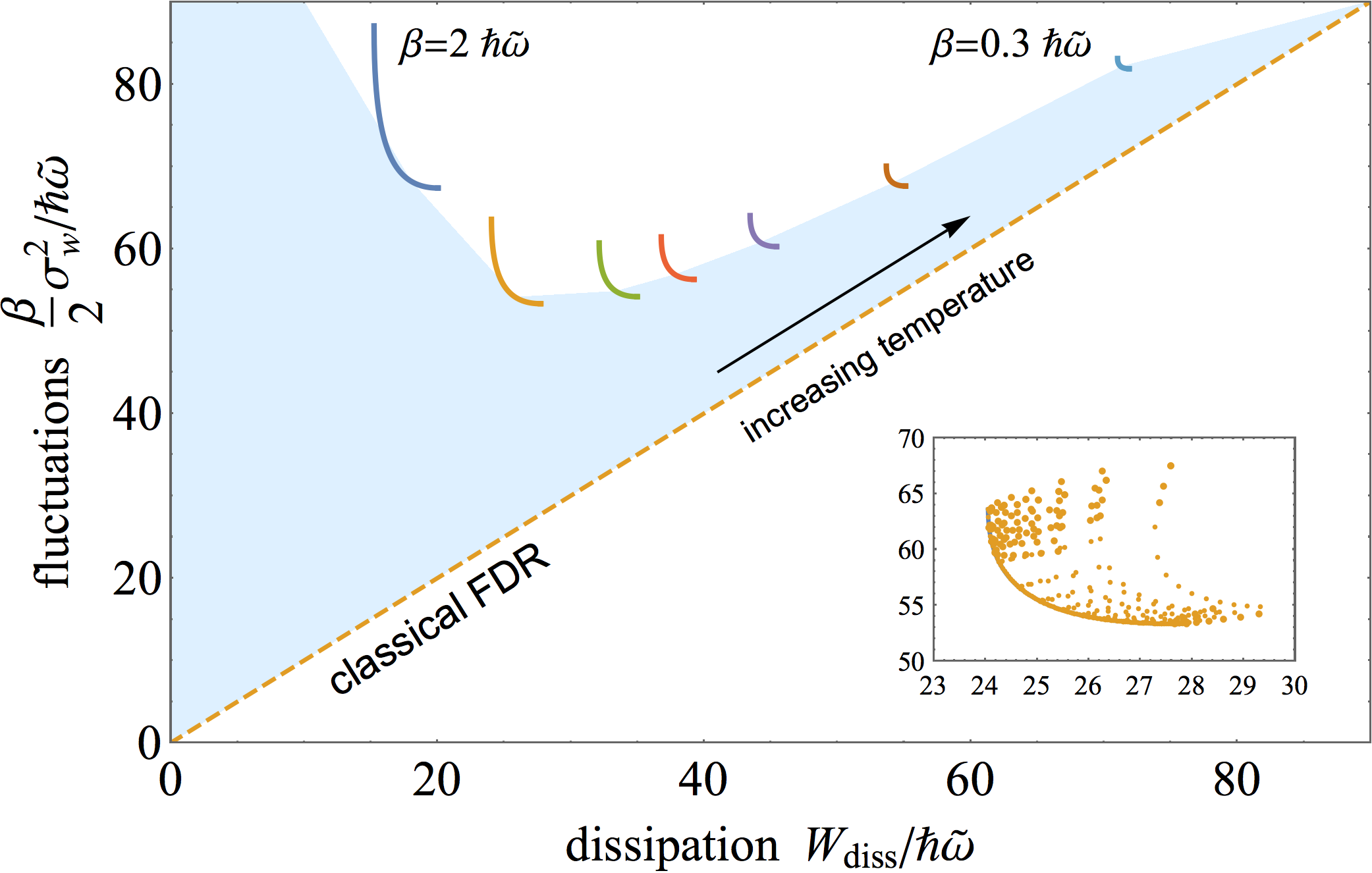}
	\caption{\blue{Pareto fronts limiting the accessible region of fluctuations $\frac{1}{2} \beta\sigma^2_w$ and dissipation $W_{\rm diss}$  for the harmonic oscillator example over all possible protocols $\{\omega_t\}$ between the end points $\omega_0 = 0.1 \tilde{\omega}$ to $\omega_1 = 10 \tilde{\omega}$ for a fixed reference frequency $\tilde{\omega}$. Curves are obtained by varying the weight $\alpha \in [0, 1]$, and for each $\alpha$ choosing the protocol to follow the geodesic that minimises $\mathcal{C}_\alpha$. Each curve is for a specific inverse temperature  $\beta=2\hbar \tilde \omega$ (blue), $\beta=1\hbar \tilde \omega$ (yellow), $\beta=0.7\hbar \tilde \omega$ (green), $\beta=0.6\hbar \tilde \omega$ (red),  $\beta=0.5\hbar \tilde \omega$ (purple),  $\beta=0.4\hbar \tilde \omega$ (brown),  and $\beta=0.3\hbar \tilde \omega$ (light blue). The blue shaded region denotes the separation between the quantum optimal protocols (Pareto fronts)  and the classical optimal protocols (diagonal) for varying $\beta$.
	Inset: Magnified Pareto front for $\beta = \hbar \tilde \omega$ and including points for suboptimal protocols, illustrating the accessible part of the fluctuation-dissipation plane.}}
	\label{fig:manypareto1}
\end{figure}

Turning to optimisation, we now use the metric $g_\alpha(\omega_t)=\alpha\Lambda(\omega_t)+(1-\alpha)\xi(\omega_t)$ associated with Eq.~\eqref{eq:alphaObjective} to  construct geodesics that interpolate between minimally dissipating and minimally fluctuating protocols (see Appendix F). So-called Pareto fronts can be used to bound the region of allowed protocols~\cite{Solon2018a}. This is illustrated in Fig.~\ref{fig:manypareto1}, where Pareto front curves indicate the trade-off between minimal fluctuation $\left(\beta \sigma^2_w/2 \right)$ and minimal dissipation ($\Wdis$) for various values of $\beta$. Each curve is obtained by evaluating $\beta \sigma^2_w/2$ and $\Wdis$ for the geodesics minimising $\mathcal{C}_\alpha$ for all values $\alpha\in[0,1]$. If the classical FDR would hold, each curve would collapse into a single point along the diagonal line $\beta\sigma^2_w/2 = \Wdis$. The quantum correction moves each Pareto front above this line and expands it from a single point to a curve, parametrised by $\alpha$. As expected, this effect is most significant at low temperatures where quantum fluctuations dominate.

\medskip

\textit{Conclusions:} In this article, we have studied the statistics of work in slowly driven open quantum systems interacting with a thermal environment. We have derived a quantum FDR for work as shown in Eq.~\eqref{eq:qFDR}, which generalises the well-known classical  FDR given by Eq.~\eqref{eq:FDR}. This result implies that whenever quantum coherence is generated during the dynamics of a slow protocol, then   $\Wdis<\frac{1}{2} \, \beta \, \sigma_{w}^{2} $, which is a genuinely quantum effect. \blue{Let us briefly comment on the generality of our results. While \eqref{eq:qFDR} has been derived using the TPM approach with thermal initial conditions, we prove in Appendix C that~\eqref{eq:qFDR} holds more generally for arbitrary initial states using alternative definitions of work based on weak measurements~\cite{Clerk2011,Bednorz2012b,Hofer2016,Hofer2017e,Allahverdyan2014c,Solinas2018,Solinas2017}. This follows directly because these measurement schemes give rise to the same work average and variance.  The validity of the quantum FDR for various work definitions highlights that the quantum effects captured by $Q_w$ stem from the coherent dynamics of the protocol, rather than arise as a result of measurement disturbance or a particular choice of work definition (see discussion in Appendix C). }

\blue{It is also interesting to discuss how breaking any of the three main assumptions used to derive the quantum FDR --namely  (i) slow driving, (ii) thermalisation, and (iii) weak coupling-- can   affect it. Both (i) and (ii) appear crucial: in Appendix H  we compare  $\Wdis $ and  $\sigma_{w}^{2}$ for a spin in contact with a bosonic bath and, while we verify the validity of Eq.~\eqref{eq:qFDR} for sufficiently slow driving, we do find violations of the FDR for faster driving. Regarding assumption (ii), one can demonstrate that the quantum FDR can break down if the system is not close to thermal equilibrium even if the dynamics are slow, as shown in \cite{zawadzki2019work} for closed unitary evolutions. On the other hand, we believe  that the quantum FDR can remain valid away from the weak coupling regime (i.e. if (iii) is broken): a step towards proving this hypothesis is done in Appendix G. By using a a discrete model of quasi-isothermal processes \cite{Gallego_2014,Perarnau_2018c}, we derive  an analogous quantum FDR for a system strongly coupled to a thermal bath.}

The quantum FDR also implies that it is fundamentally impossible to simultaneously minimise dissipation and fluctuations in slow coherent quantum processes. In the second part of the paper  we have derived a family of metrics whose geodesics interpolate between minimally-dissipative and minimally-fluctuating thermodynamic protocols, and our results unveil a new geometric structure within quantum thermodynamics.  A promising platform to observe these effects experimentally are quantum dots~\cite{Pekola2015,Ludovico2016,alonso2018work} and superconducting qubits \cite{PekolaMaxwell2016,Cottet2018},  where slowly driven non-commuting protocols appear as a realistic possibility~\cite{HacohenGourgy2016} \blue{and proposals for observing TPM work statistics using a calorimeter have been made \cite{Pekola2013b}}.  An interesting future direction is \textcolor{black}{to extend the FDR to many-body closed systems \cite{Fusco2014,zawadzki2019work,arrais2019work}}, and to investigate how these genuinely quantum effects can modify the thermodynamic uncertainty relations in non-equilibrium steady states~\cite{Barato2015,Ptaszy2018,Agarwalla2018,guarnieri2019thermodynamics} and FDR's in other contexts such as quantum transport~\cite{Averin2010}. 

\medskip

\textit{Acknowledgements.} We thank Sreekanth Manikandan, Irene D'Amico and Augusto Roncaglia for interesting discussions and comments on the manuscript. \blue{This research was supported in part by the COST network MP1209 ``Thermodynamics in the quantum regime'' and by the National Science Foundation under Grant No. NSF PHY-1748958. J.A. acknowledges support from EPSRC (grant EP/R045577/1) and the Royal Society.}

\bibliographystyle{apsrev4-1}
\bibliography{mybib.bib}

\appendix

\widetext

\section{Work moments for open system dynamics}\label{app:MGF}

\noindent 

\blue{In the main text we define the fluctuating work done on the system using the Two-Projective-Measurement (TPM) scheme. Let us denote the spectral decomposition of the total system-bath Hamiltonian $\SB{H}(t)=\sum_n \epsilon_n (t)\ket{\epsilon_n(t)}\bra{\epsilon_n(t)}$ and define the time-ordered unitary $U(t_f,t_i)=\timeorderedexp\big(-(i/\hbar)\int^{t_f}_{t_i} dt' \ \SB{H}(t')\big)$ generated by variations in the global Hamiltonian. The TPM work distribution $P(w)$ is constructed in terms of observed transitions between energy states \cite{Talkner2007c}, resulting in 
\begin{align}
P(w)=\sum_{n,m}\delta[w-\epsilon_m(\tau)+\epsilon_n(0)] p_{m | n} \ p_n,
\end{align}
where $p_n=\bra{\epsilon_n(0)}\SB{\pi}(0)\ket{\epsilon_n(0)}$ is the initial energy occupation probability while $p_{m | n}=\big|\bra{\epsilon_m(\tau)}U(\tau,0)\ket{\epsilon_n(0)}\big|^2$ denotes the conditional energy transition probability.} After taking a Fourier transform, the expression for the corresponding moment generating function of the work is given by:
\begin{align}\label{eq:mgfStart}
G(\lambda) &= \average{e^{i\lambda w}}{}=   \TrSB{U^{\dagger}(\tau,0) e^{i\lambda \SB{H}(\tau)} U(\tau,0) e^{-i\lambda \SB{H}(0)} \SB{\pi}(0)}.
\end{align}
Note that this function implicitly assumes that $\SB{\pi}(0)$ commutes with the initial Hamiltonian, which is guaranteed since $\SB{\pi}(0)=e^{-\beta \SB{H}(0)}/\TrSB{e^{-\beta \SB{H}(0)}}$ is a global thermal state with negligible coupling between system and bath. From~\eqref{eq:mgfStart} one can compute the first two work moments using $\langle w^k \rangle=(-i)^{k} (d^k/d\lambda^k)G(\lambda)\big|_{\lambda=0}$, and a lengthy but straightforward calculation yields the following \cite{Suomela2014}:
\begin{align}
\label{eq:av}&\langle w\rangle=\int^\tau_0 dt \ \TrS{\sys{\dot{H}}(t)\sys{\varrho}(t)}, \\
&\langle w^2\rangle= 2 \Real \int^\tau_0 dt_{1} \int^{t_{1}}_0 dt_{2} \ \TrSB{\sys{\dot{H}}^H(t_{1}) \sys{\dot{H}}^H(t_{2})\SB{\pi}(0)}. \label{eq:moment2}
\end{align}
where we denote \blue{$X^H_{\scriptscriptstyle S}(t)=U^\dagger(t,0) \, X_{\scriptscriptstyle S}(t)\, U(t,0)$ as operator $\sys{X}(t)$} in the Heisenberg picture. The expressions~\eqref{eq:av} and ~\eqref{eq:moment2} are valid for any global unitary evolution \blue{driven by a total Hamiltonian $\SB{H}(t)=\sys{H}(t)+\bath{H}+\SB{V}$ with time-dependence only on the system Hamiltonian, and thus $\SB{\dot{H}}(t)=\sys{\dot{H}}(t)$. Note that for $\langle w^2\rangle$ in~\eqref{eq:moment2} the Heisenberg picture $ \sys{\dot{H}}^H(t)$ is used instead of $\sys{\dot{H}}(t)$, and thus the trace is taken over the whole Hilbert space due to the bath-dependence of the unitary $U(t,0)$.} 

Our goal will be to express the second moment in terms of the reduced density operator $\sys{\varrho}(t)=\TrB{\SB{\varrho}(t)}$, where we define the evolved density operator for the composite state by $\SB{\varrho}(t)=U(t,0)\SB{\pi}(0)U^\dagger(t,0)$. To do this we now assume that the evolution of the system is of Lindblad form as defined in the main text:
\begin{align}\label{eq:lindblad}
\sys{\dot{\varrho}}(t):=-\frac{i}{\hbar}\TrB{[\SB{H}(t),\SB{\varrho}(t)]}=\mathscr{L}_t[\sys{\varrho}(t)],
\end{align}
with $\mathscr{L}_t[(.)]$ a time-dependent Lindbladian. Implicit within our assumption for~\eqref{eq:lindblad} is the Born-Markov approximation, which assumes that the global state remains factorised at all times during the evolution \cite{breuer2002theory}:
\begin{align}\label{eq:born}
\SB{\varrho}(t)\simeq\sys{\varrho}(t)\otimes \bath{\pi},
\end{align}
This assumption is justified only in the weak-coupling regime.

Our goal will now be to use~\eqref{eq:lindblad} to rewrite~\eqref{eq:moment2} in terms of the system degrees of freedom. Let us now consider two hermitian \blue{time-dependent} operators \blue{$A_{\scriptscriptstyle S}(t), B_{\scriptscriptstyle S}(t)$} acting on the system Hilbert space alone. We are concerned with evaluating the two-time correlation function  $\langle A^H_{\scriptscriptstyle S}(t') \, B^H_{\scriptscriptstyle S}(t) \rangle $ \blue{in Heisenberg picture} for  $t'\geq t$, which can be expressed as follows:
\begin{align}
\nonumber \langle A^H_{\scriptscriptstyle S}(t') \, B^H_{\scriptscriptstyle S}(t) \rangle
&=\TrSB{A^H_{\scriptscriptstyle S}(t') B^H_{\scriptscriptstyle S}(t)\SB{\pi}(0)}, \\
\nonumber &=\TrSB{U^\dagger(t',0) \, A_{\scriptscriptstyle S} \blue{(t')} \, U(t',0) \, U^\dagger(t,0) \, B_{\scriptscriptstyle S} \blue{(t)} \, U(t,0) \, \SB{\pi}(0)}, \\
\nonumber &=\TrSB{A_{\scriptscriptstyle S}\blue{(t')} \, U(t',t) \, B_{\scriptscriptstyle S} \blue{(t)} \, \SB{\varrho}(t) \, U^\dagger(t',t)}, \\
&=\TrS{A_{\scriptscriptstyle S}\blue{(t')} \, \TrB{U(t',t) \, B_{\scriptscriptstyle S} \blue{(t)} \, \SB{\varrho}(t)U^\dagger(t',t)}},
\end{align}
where in the third line we used $\SB{\pi}(0)=U^\dagger(t,0)\SB{\varrho}(t)U(t,0)$ and the cyclicity of the trace, while in the final line we used the fact that $B_{\scriptscriptstyle S} \blue{(t)}=B_{\scriptscriptstyle S} \blue{(t)} \otimes \bath{\mathbb{I}}$. Setting $\tilde{t}=t'-t\geq 0$, a simple change in variables gives
\begin{align}\label{eq:correlation}
\langle A^H_{\scriptscriptstyle S}(t+\tilde{t}) B^H_{\scriptscriptstyle S}(t) \rangle=\TrS{A_{\scriptscriptstyle S} \blue{(t+\tilde{t})}  \ \chi_{\scriptscriptstyle S}(\tilde{t})}
\end{align}
where 
\begin{align}
\left\{\begin{array}{l}
\SB{\chi}(\tilde{t})=U(t+\tilde{t},t) \, B_{\scriptscriptstyle S}\blue{(t)} \, \SB{\varrho}(t)U^\dagger(t+\tilde{t},t), \\
\chi_{\scriptscriptstyle S}(\tilde{t})=\TrB{\SB{\chi}(\tilde{t})}.
\end{array}\right.
\end{align}
Now observe that $\SB{\chi}(\tilde{t})$ is the solution to the following equation of motion:
\begin{align}\label{eq:LVN}
\frac{d}{d\tilde{t}}\SB{\chi}(\tilde{t}):=-\frac{i}{\hbar}[\SB{H}(\tilde{t}),\SB{\chi}(\tilde{t})],
\end{align}
with initial condition $\SB{\chi}(0)=B_{\scriptscriptstyle S} \blue{(t)} \, \SB{\varrho}(t)$. We now use the Born-Markov approximation~\eqref{eq:born}, which implies that initial condition to~\eqref{eq:LVN} factorises according to $\SB{\chi}(0)=B_{\scriptscriptstyle S}  \blue{(t)}  \sys{\rho}(t)\otimes \bath{\pi}$. Given that the initial operator $\SB{\chi}(0)$ here factorises and obeys the same global equation of motion given by~\eqref{eq:LVN} with respect to $\tilde{t}$ as the state $\SB{\varrho}(t)$, we obtain the following solution after tracing out the bath degrees of freedom:
\begin{align}
\chi_{\scriptscriptstyle S}(\tilde{t})=\overleftarrow{P}(t+\tilde{t},t)[B_{\scriptscriptstyle S} \blue{(t)} \, \varrho_{\scriptscriptstyle S}(t)],
\end{align}
where 
\begin{align}
\overleftarrow{P}(t_1,t_2)=\overleftarrow{\mathcal{T}}\text{exp}\bigg(\int_{t_2}^{t_1}d\nu \ \mathscr{L}_\nu \bigg),
\end{align}
is the propagator for the Lindbladian in~\eqref{eq:lindblad}. Combining this with~\eqref{eq:correlation} we have
\begin{align}\label{eq:correlation2}
\langle A^H_{\scriptscriptstyle S}(t+\tilde{t}) B^H_{\scriptscriptstyle S}(t) \rangle=\TrS{A_{\scriptscriptstyle S} \ \overleftarrow{P}(t+\tilde{t},t)[B_{\scriptscriptstyle S} \blue{(t)}  \,\varrho_{\scriptscriptstyle S}(t)]}.
\end{align}
Setting $\sys{A} \blue{(t_1)} =\sys{\dot{H}}(t_{1})$ and $\sys{B} \blue{(t_2)} =\sys{\dot{H}}(t_{2})$, and combining this all together gives us an expression for $\langle w^2\rangle$ from~\eqref{eq:moment2} in terms of the system degrees of freedom:
\begin{align}\label{eq:moments4}
\langle w^2 \rangle=2 \Real \int^\tau_0 dt_{1} \int^{t_{1}}_0 dt_{2} \ \TrS{\sys{\dot{H}}(t_{1}) \ \overleftarrow{P}(t_{1},t_{2})[\sys{\dot{H}}(t_{2})\sys{\varrho}(t_{2})]}.
\end{align}
Furthermore, the squared average work~\eqref{eq:av} can be written as follows:
\begin{align}\label{eq:average2}
\nonumber \langle w \rangle^2&=2 \Real \int^\tau_0 dt_{1} \int^{t_{1}}_0 dt_{2} \ \TrS{\sys{\dot{H}}(t_{1})\sys{\varrho}(t_{1})}\TrS{\sys{\dot{H}}(t_{2})\sys{\varrho}(t_{2})}, \\
&=2 \Real \int^\tau_0 dt_{1} \int^{t_{1}}_0 dt_{2} \ \TrS{\sys{\dot{H}}(t_{1}) \ \overleftarrow{P}(t_{1},t_{2})[\TrS{\sys{\dot{H}}(t_{2})\sys{\varrho}(t_{2})} \ \sys{\varrho}(t_{2})]}.
\end{align}
We now define $\Delta_\varrho A=A-\Tr{A\varrho}$ and combine~\eqref{eq:moments4} and~\eqref{eq:average2} to get
\begin{align}
\nonumber\sigma_w^2&=\langle w^2 \rangle-\langle w \rangle^2 \\
\nonumber&=2 \Real \int^\tau_0 dt_{1} \int^{t_{1}}_0 dt_{2} \ \TrS{\sys{\dot{H}}(t_{1}) \ \overleftarrow{P}(t_{1},t_{2})[ \Delta_{\sys{\varrho}(t_{2})}\sys{\dot{H}}(t_{2})\sys{\varrho}(t_{2})]}, \\
\nonumber&= \int^\tau_0 dt_{1} \int^{t_{1}}_0 dt_{2} \ \TrS{\sys{\dot{H}}(t_{1}) \ \overleftarrow{P}(t_{1},t_{2})[ \curbra{\Delta_{\sys{\varrho}(t_{2})}\sys{\dot{H}}(t_{2}),\sys{\varrho}(t_{2})}}, \\
&=2\int^\tau_0 dt_{1} \int^{t_{1}}_0 dt_{2} \ \TrS{\sys{\dot{H}}(t_{1}) \ \overleftarrow{P}(t_{1},t_{2})[ \operatorS_{\sys{\varrho}(t_{2})}(\sys{\dot{H}}(t_{2}))},
\end{align}
where in the third line we used the fact that $\Real \ \Tr{A \overleftarrow{P}(t_{1},t_{2})[B]}=(1/2)\Tr{A \overleftarrow{P}(t_{1},t_{2})[B+B^{\dagger}]}$ for $A=A^\dagger$, and in the fourth we introduced the definition for $\operatorS_{\varrho}$ from Eq. (3). This concludes the derivation of Eq. (2) in the main text.

\section{Derivation of Eq. (4)}\label{app:quasistaticFluctuations}
\noindent We want to take the slow driving limit of the expression:
\begin{align}\label{eq:D1}
\sigma^2_{ w}=2 \int^\tau_0 dt_{1} \int^{t_{1}}_0 dt_{2} \ \Tr{\dot{H}_{t_{1}}\overleftarrow{P}(t_{1},t_{2})\big(\operatorS_{\varrho_{t_{2}}}(\dot{H}_{t_{2}})\big)}.
\end{align}
Recalling the definition of $\overleftarrow{P}(t_{1},t_{2})=\timeorderedexp\big(\int_{t_{2}}^{t_{1}} d\nu \mathscr{L}_\nu \big)$, we notice that the trace will decay to zero exponentially fast in \newline $|t_{1}-t_{2}|\sim \tau$, since $\operatorS_{\varrho_{t_{2}}}(\dot{H}_{t_{2}})$ is traceless. For this reason, we can substitute at first order in $1/\tau$ the varying Liouvillian with the initial one:
\begin{align}
\label{eq:second_prel}
\sigma^2_{ w}&\simeq 2\int_{0}^{\tau}dt_{1}\int_{0}^{t_{1}}dt_{2}\,\Tr{\dot H_{t_{1}}\, e^{(t_{1}-t_{2})\lind_{t_{2}}} \sqrbra{\operatorS_{\varrho_{t_{2}}}(\dot{H}_{t_{2}})}},\\
\nonumber&= 2\int_{0}^{\tau}dt_{1}\int_{0}^{t_{1}}ds\,\Tr{\dot H_{t_{1}}\,e^{s \lind_{t_{1}-s}} \sqrbra{\operatorS_{\varrho_{t_1-s}}(\dot{H}_{t_1-s})}},
\end{align}
where in the second line we made the substitution $s = t_{1}-t_{2}$. Again, since $s$ will be typically much bigger than the thermalisation timescales, not only we can approximate $t_{1}-s$ with $t_{1}$ in all the expression (since the correction for finite $s$ will be exponentially suppressed), but also we can send the limit of the integration to infinity. Then, equation \blue{\eqref{eq:second_prel}} becomes:
\begin{align}
 \nonumber\sigma^2_{ w} &\simeq 2\int_{0}^{\tau}dt_{1}\int_{0}^{t_{1}}ds\,\Tr{\dot H_{t_{1}}\,e^{s \lind_{t_{1}}} \sqrbra{\operatorS_{\varrho_{t_{1}}}(\dot{H}_{t_{1}})}},\\
 \nonumber&\simeq 2\int_{0}^{\tau}dt_{1}\int_{0}^{\infty}ds\,\Tr{\dot H_{t_{1}}\,e^{s \lind_{t_{1}}} \sqrbra{\operatorS_{\varrho_{t_{1}}}(\dot{H}_{t_{1}})}}, \\
&= -2\int_{0}^{\tau}dt\,\Tr{\dot H_{t}\lind^{+}_{t} \sqrbra{\operatorS_{\varrho_{t}}(\dot{H}_{t})}},
\end{align}
where in the last step we used the integral expression of the Drazin inverse $\lind_t^{+}$ in Eq. (5) and the fact that $\operatorS_{\varrho_{t_{1}}}(\dot{H}_{t_{1}})$ is traceless. Finally, at first order in $1/\tau$, we can substitute $\varrho_t\simeq\pi_t$. This concludes the derivation of Eq. (4) in the main text.

 \section{Weak measurements of fluctuating work}

\blue{Our analysis throughout the paper defines work via the standard two-projective measurement (TPM) scheme. This definition of work is typically adopted for states that are initially thermal, as one recovers the usual Jarzynski equality and laws of thermodynamics at the ensemble level \cite{Talkner2009} while remaining consistent with the usual definition of work in the classical regime \cite{Jarzynski2015}. Furthermore, since the initial state is diagonal in the energy basis one can neglect the effect of measurement backaction caused by the initial projective energy measurement. However, for states that are non-diagonal the TPM scheme can remove initial coherences due to disturbances caused by the first projective measurement. As a result, one can no longer associate the average TPM work to the change in total energy of the system and bath \cite{baumer2018fluctuating}.}

\blue{In order to characterise work when initial coherences are present, alternative definitions of work based around weak measurement have been proposed that preserve the coherent evolution of the system and bath \cite{Allahverdyan2014c,Solinas2018,Solinas2017}. In these measurement schemes one may obtain negative quasi-probabilities in the work distribution, signifying uniquely quantum behaviour such as contextuality \cite{Lostaglio2018}. In this Appendix we will demonstrate that definitions of work based on weak measurement gives rise to the same fluctuation-dissipation relation Eq. (8), but are now applicable to arbitrary initial states of the system that may be non-diagonal in the energy basis. While different choices of measurement scheme typically lead to different work statistics, any discrepancies between different definitions of quantum work only apply to the moments of third order and higher, which are not relevant to the quantum work FDR Eq. (8). We will therefore demonstrate that our results hold more generally beyond the TPM definition of work.}

\blue{We first describe a continuous weak measurement scheme that can be implemented by coupling only to the system degrees of freedom. Recall that in the absence of any measurement, the system and bath evolve according to state $\SB{\varrho}(t)=U(t,0)\SB{\varrho}(0)U^\dagger(t,0)$, with unitary $U(t_f,t_i)=\timeorderedexp\big(-i\int^{t_f}_{t_i} dt' \ \SB{H}(t')\big)$ and setting $\hbar=1$. In contrast to the main text, we now assume arbitrary initial conditions for the system state, such that $\SB{\varrho}(0)=\sys{\rho}(0)\otimes \bath{\pi}$. Here $\sys{\rho}(0)$ may not be thermal or even commute with its Hamiltonian, ie. $[\sys{\rho}(0),\sys{H}(0)]\neq 0$.}

\blue{In this weak measurement approach, the fluctuations in work can be characterised as the time-integrated fluctuations in the power observable $\sys{\dot{H}}(t)$ \cite{Hofer2017e}. To determine these fluctuations we couple the system and bath to a detector modelled as a two-level system, initially uncorrelated such that $\SBD{\varrho}(t)=\SB{\varrho}(0)\otimes \De{\varrho}(0)$. The modified Hamiltonian including the interaction with the detector is now given by 
\begin{align}
\SBD{H}(t)=\SB{H}(t)+\frac{\lambda}{2}\sys{\dot{H}}(t)\otimes\sigma_z,
\end{align}
where $\lambda$ is a coupling constant and $\sigma_z$ is the Pauli spin-z operator. This generates a new evolution \newline $\tilde{U}(t,0)=\timeorderedexp\big(-i\int^{t}_{0} dt' \ \SBD{H}(t)\big)$, and we denote the reduced state of the detector as a function of $\lambda$ by $\De{\varrho}(t;\lambda)$. By measuring the relative change in phase of the detector  one gets the so-called \textit{Keldysh-ordered} full-counting statistics \cite{Clerk2011,Bednorz2012b,Hofer2016,Hofer2017e}:
\begin{align}\label{eq:FCS}
\tilde{G}(\lambda):=\frac{\bra{\uparrow}\De{\varrho}(\tau;\lambda)\ket{\downarrow}}{\bra{\uparrow}\De{\varrho}(0;\lambda)\ket{\downarrow}}=\TrSB{ V_{\lambda/2}(\tau)\SB{\varrho}(0) V^\dagger_{-\lambda/2}(\tau)},
\end{align}
where
\begin{align}
V_\lambda(t)=\timeorderedexp\big(i\lambda\int^{t}_{0} dt' \ \sys{\dot{H}}^H(t')\big).
\end{align}
where $\sys{\dot{H}}^H(t')$ is in the Heisenberg picture with respect to the isolated system and bath unitary dynamics. In this approach one interprets~\eqref{eq:FCS} as a moment generating function for the fluctuating work, with moments
\begin{align}\label{eq:FCSmoments3}
\langle w^k \rangle=(-i)^{k} (d^k/d\lambda^k)\tilde{G}(\lambda)\big|_{\lambda=0}.
\end{align}
One can see that these statistics generally differ from those obtained via the TPM scheme via~\eqref{eq:mgfStart}, as highlighted in \cite{Hofer2017e}. However, we will now prove that the first two moments of~\eqref{eq:FCS} are in fact equivalent to those obtained from~\eqref{eq:mgfStart} used in the main text, but valid for any initial system state. }

\blue{We first use the Magnus expansion for $V_\lambda(t)=\exp{(\Omega_\lambda(t))}$, where
\begin{align}\label{eq:magnus}
\Omega_\lambda(t)=i\lambda\int^t_0 dt_1 \ \sys{\dot{H}}^H(t_1)+\frac{\lambda^2}{2}\int^t_0 dt_1 \int^{t_1}_0 dt_2 \ [\sys{\dot{H}}^H(t_1),\sys{\dot{H}}^H(t_2)]+\mathcal{O}(\lambda^3),
\end{align}
The generating function~\eqref{eq:FCS} can then be expressed in powers of $\lambda$:
\begin{align}\label{eq:FCS2}
\tilde{G}(\lambda)=1+i\lambda\int^\tau_0 dt \ \TrS{\sys{\dot{H}}(t)\sys{\varrho}(t)}+ \lambda^2 \Real \int^\tau_0 dt_{1} \int^{t_{1}}_0 dt_{2} \ \TrSB{\sys{\dot{H}}^H(t_{1}) \sys{\dot{H}}^H(t_{2})\SB{\varrho}(0)}+\mathcal{O}(\lambda^3),
\end{align}
It should be noted here that the second order term in~\eqref{eq:magnus} is skew hermitian and hence does not contribute to the second order term in~\eqref{eq:FCS2}. Finally using the definition of the work moments, we get
\begin{align}\label{eq:workFCS1}
&\langle w\rangle=\int^\tau_0 dt \ \TrS{\sys{\dot{H}}(t)\sys{\varrho}(t)}, \\
&\langle w^2\rangle= 2 \Real \int^\tau_0 dt_{1} \int^{t_{1}}_0 dt_{2} \ \TrSB{\sys{\dot{H}}^H(t_{1}) \sys{\dot{H}}^H(t_{2})\SB{\varrho}(0)}\label{eq:workFCS11}. 
\end{align}
Comparison with~\eqref{eq:av} and~\eqref{eq:moment2} confirms that these first two moments of work~\eqref{eq:workFCS1} and~\eqref{eq:workFCS11} take the same form as the TPM definition used in the main text, but now hold for any initial conditions chosen for the system. With these expressions the proof of the quantum work FDR follows exactly as before for any initial state.  }

\blue{As an alternative to continuous weak measurement, full-counting statistics for work can be obtained through discrete coupling to the Hamiltonian. This gives rise to a different moment generating function \cite{Solinas2018}:
\begin{align}\label{eq:FCSdiscrete}
\tilde{G}(\lambda)=\TrSB{U^{\dagger}(\tau,0) e^{i\lambda \SB{H}(\tau)} U(\tau,0) e^{-i\lambda \SB{H}(0)} \SB{\varrho}(0)}.
\end{align}
It is straightforward to show that the first and second moments here become
\begin{align}\label{eq:workFCS2}
&\langle w\rangle=\TrSB{\big(U^\dagger(\tau,0)\SB{H}(\tau)U(\tau,0)-\SB{H}(0)\big)\SB{\rho}(0)}, \\
&\langle w^2\rangle=\TrSB{\big(U^\dagger(\tau,0)\SB{H}(\tau)U(\tau,0)-\SB{H}(0)\big)^2\SB{\rho}(0)}.
\end{align}
As shown in \cite{Suomela2014}, these expressions are exactly equivalent to the time-integrated expressions~\eqref{eq:workFCS1} and~\eqref{eq:workFCS11} since we assume only the system Hamiltonian to be time-dependent. We remark that another definition of work based on the Margenau-Hill quasi-probability distribution has been proposed by Allahverdyan in \cite{Allahverdyan2014c}, which is obtained from an alternative two-point measurement scheme combining both strong and weak energy measurements. Finally, a quasi-probability for work constructed within the quantum histories approach is proposed in \cite{Miller2017}. Again, in these cases one finds the same first two moments of work~\eqref{eq:workFCS2}, which equate to~\eqref{eq:workFCS1} and~\eqref{eq:workFCS11}. We reiterate that all of these quasi-probabilistic generalisations of the TPM work distribution for arbitrary initial states differ only for third moments and higher, and therefore the particular choice is not relevant for deriving the quantum work FDR, which is a statement only about the first and second work moments.}

\blue{In conclusion, we have shown that the quantum work FDR Eq. (8) continues to hold for arbitrary initial states using alternative definitions of quantum work based on weak measurement. Therefore Eq. (8) is not only restricted to the TPM definition of work, or a particular choice of initial conditions. This implies that the quantum modifications to the work FDR are a manifestation of the coherent dynamics generated during the slow driving protocol, and are not a signature of quantum measurement effects stemming from a particular choice of measurement scheme. Furthermore, we have presented a continuous weak measurement scheme that can verify Eq. (8) using only local interactions with the system degrees of freedom, as opposed to measurements of the full system-bath Hilbert space. }

\medskip

\section{Proof of Eq. (9)}\label{app:positivity}

In this Appendix we prove the positivity of the dynamical skew information $\mathcal{I}_t(\pi_t,\dot{H}_t)$, which under time integration gives the quantum correction $Q_w=\beta\int^\tau_0 dt \ \mathcal{I}_t(\pi_t,\dot{H}_t)$ appearing in Eq. (8). Consider the Hilbert space $M_d$ of $d\times d$ complex matrices with Hilbert-Schmidt inner product $\langle A, B \rangle=\Tr{B^\dagger A}$. Then any superoperator $\mathcal{M}(.)$ acting on the elements of this Hilbert space can be expressed as a $d^2 \times d^2$ matrix. The matrix describing $\mathcal{M}(.)$ is positive if $\Tr{A^\dagger \mathcal{M}(A)}\geq 0$ for any $A\in M_d$, and we define the adjoint $\mathcal{M}^\dagger(.)$ as the superoperator satisfying $\Tr{\mathcal{M}^\dagger(B^\dagger) A}=\Tr{B^\dagger \mathcal{M}(A)}$. We begin by assuming a generic interaction between system and bath formed by a sum of hermitian operators
\begin{align}
\SB{V}=\sum_\alpha A_\alpha \otimes B_\alpha.
\end{align}
As stated in the main text, we will work in the slow driving regime and assume that the bath dynamics are much faster than the driving rate of the system Hamiltonian. This means that one can neglect any non-adiabatic contributions to the reduced system dynamics \cite{Yamaguchi2017,Dann}. In addition the system dynamics are assumed to satisfy detailed balance along with the Born-Markov and rotating-wave approximations \cite{breuer2002theory,alicki2007quantum}. When taken together these assumptions result in a time-dependent Markovian master equation describing the system dynamics that can be expressed in a Lindblad form $\mathscr{L}_t(.)=-i[H_t,(.)]+\mathcal{D}_t(.)$, and a precise derivation of its structure and regime of validity can be found in \cite{Albash2012}. Throughout this derivation we will only be concerned with the structure of the Lindbladian at some fixed point in time. At any time $t$ the time-dependent Lindbladian takes the following form \cite{Albash2012,Yamaguchi2017,Dann}:
\begin{align}
\nonumber&\mathcal{U}_t(.)=-i[H_t,(.)], \\
&\mathcal{D}_t(.)=\sum_{\omega_t} \sum_{\alpha,\beta} \gamma_{\alpha\beta}(\omega_t)\bigg(A_\beta(\omega_t)(.)A^\dagger_\alpha (\omega_t)-\frac{1}{2}\lbrace A^\dagger_\alpha (\omega_t)A_\beta(\omega_t),(.)\rbrace\bigg),
\end{align}
where $\gamma_{\alpha\beta}(\omega_t)$ is a hermitian matrix representing the Fourier transform of the bath correlation function, and we have defined the eigenoperators
\begin{align}
A_\alpha(\omega_t)=\sum_{\omega_t=\epsilon_j(t)-\epsilon_i(t)}\ket{\epsilon_i(t)}\bra{\epsilon_i(t)} A_\alpha \ket{\epsilon_j(t)}\bra{\epsilon_j(t)},
\end{align}
with $H_t=\sum_j \epsilon_j(t) \ket{\epsilon_j(t)}\bra{\epsilon_j(t)}$ the spectral decomposition of the system Hamiltonian at some fixed point in time. The eigenoperators satisfy
\begin{align}\label{eq:Aw}
A^\dagger_\alpha(\omega_t)=A_\alpha(-\omega_t).
\end{align}
It then follows that such a Lindbladian has a unique zero eigenvalue corresponding to a thermal fixed point $\mathscr{L}_t(\pi_t)=0$, while all other eigenvalues have a strictly negative real part \cite{Rivas2011}. This ensures that at each fixed configuration $\mathscr{L}_t$, any initial state $\rho$ will converge to the instantaneous fixed point:
\begin{align}
\lim_{\nu\to\infty}e^{\nu\mathscr{L}_t}(\rho)=\pi_t.
\end{align}
We now observe some important properties of $\mathscr{L}_t(.)$ \cite{breuer2002theory}. Firstly, due to the rotating-wave approximation the unitary and dissipative parts commute:
\begin{align}\label{eq:RW}
[\mathcal{U}_t(.),\mathcal{D}_t(.)]=[\mathcal{U}_t(.),\mathcal{D}_t^\dagger(.)]=0.
\end{align}
Secondly, $\mathscr{L}_t(.)$ satisfies the condition of \textit{detailed balance}, which implies
\begin{align}\label{eq:DB}
\nonumber&\pi_t \ A_\alpha (\omega_t)=e^{\beta \omega_t}A_\alpha(\omega_t)\pi_t, \\
&\pi_t \ A^\dagger_\alpha (\omega_t)=e^{-\beta \omega_t}A^\dagger_\alpha (\omega_t)\pi_t.
\end{align}
Finally, the bath correlation function satisfies the KMS condition and hence
\begin{align}\label{eq:KMS}
\gamma_{\alpha \beta}(-\omega_t)=e^{-\beta \omega_t} \gamma_{\beta\alpha}(\omega_t).
\end{align}
Now note that the dynamical skew information is a real-valued trace functional, thus it is sufficient to prove positivity of the quantity 
\begin{align}\label{eq:skew3}
\mathscr{I}(\pi_t,A):=-\text{Re} \ \Tr{A \mathscr{L}_t^+ \ \mathcal{M}_t(A)},
\end{align}
where $A=A^\dagger$ is an arbitrary hermitian operator, $\mathscr{L}_t^+$ the Drazin inverse of the Lindbladian defined in Eq. (5) and 
\begin{align}
\mathcal{M}_t(.):=\frac{1}{2}\lbrace \pi_t, (.) \rbrace-\int^1_0 ds \ \pi_t^s (.) \pi_t^{1-s}.
\end{align}
Here $\mathcal{M}(.)$ represents the difference between the arithmetic and logarithmic matrix means, and is hence a positive superoperator due to the Kubo-Ando inequality \cite{Kubo1980}. Alternatively, we can see this by looking at the spectrum of $\mathcal{M}_t(.)$. The eigenvectors are given by the energy state elements $\ket{\epsilon_i(t)}\bra{\epsilon_j(t)}$, and one finds
\begin{align}
\mathcal{M}_t(\ket{\epsilon_i(t)}\bra{\epsilon_j(t)})=\lambda_{ij}(t)\ket{\epsilon_i(t)}\bra{\epsilon_j(t)},
\end{align}
where
\begin{align}\label{eq:spectrum}
\lambda_{ij}(t)=\begin{cases} \frac{p_i(t)+p_j(t)}{2}-\frac{p_i(t)-p_j(t)}{\ln p_i(t) -\ln p_j(t) }> 0; \ \ \ \epsilon_i(t)\neq \epsilon_j(t),\\
0; \ \ \ \ \ \ \ \ \ \ \ \ \ \ \ \ \ \ \ \ \ \ \ \ \ \ \ \  \ \ \ \ \ \ \ \ \ \ \ \ \ \ \ \   \ \ \ \    \epsilon_i(t)= \epsilon_j(t).
\end{cases}
\end{align}
and $p_i(t)$ represent the eigenvalues of $\pi_t$. In addition, since $\pi_t$ commutes with Hamiltonian $H_t$ one can verify the commutation relation
\begin{align}\label{eq:comm2}
[\mathcal{M}_t(.),\mathcal{U}_t(.)]=0.
\end{align}
Let us now consider the relation between $\mathcal{M}_t(.)$ and the dissipator $\mathcal{D}_t(.)$. It is first useful to introduce the following integral representation for the matrix power $\pi^s$ for positive $\pi$ \cite{Briet2009}: 
\begin{align}
\pi^s=\int^\infty_0 d\mu_s(x) \big(e^{-x \pi} -\mathbb{I}\big); \ \ s\in(0,1).
\end{align}
with $\mu_s(x)$ a positive measure on $(0,\infty)$ that we leave unspecified for convenience. Using detailed balance~\eqref{eq:DB} we get the following:
\begin{align}\label{eq:DB2}
\nonumber\pi_t^s A_\alpha(\omega_t)&=\int^\infty_0 d\mu_s(x) \big(e^{-x \pi_t} -\mathbb{I}\big)A_\alpha(\omega_t), \\
\nonumber&=\int^\infty_0 d\mu_s(x) \bigg(\sum^\infty_{n=0}\frac{(-x)^n \, \blue{\pi_t^n} }{n!} -\mathbb{I}\bigg)A_\alpha(\omega_t), \\
\nonumber&=A_\alpha(\omega_t)\int^\infty_0 d\mu_s(x) \bigg(\sum^\infty_{n=0}\frac{(-x)^n \, \blue{(e^{\beta\omega_t}\pi_t)^n} }{n!} -\mathbb{I}\bigg), \\
&=e^{s\beta \omega_t}A_\alpha(\omega_t)\pi_t^s.
\end{align}
Similarly one finds 
\begin{align}
\pi_t^s A_\alpha^\dagger(\omega_t)=e^{-s\beta\omega_t}A_\alpha^\dagger(\omega_t) \pi_t^s.
\end{align}
This then implies
\begin{align}\label{eq:DB3}
A_\beta(\omega_t) \,  \mathcal{M}_t(.)  \, A^\dagger_\alpha(\omega_t)=e^{-\beta \omega_t} \blue{\mathcal{M}_t \,  \left[ A_\beta(\omega_t)(.)A^\dagger_\alpha(\omega_t)\right]}
\end{align}
Using this one obtains the following:
\begin{align}\label{eq:comm3}
\nonumber\mathcal{D}_t \, 
\blue{\left[\mathcal{M}_t(.)\right]}
&=\sum_{\omega_t} \sum_{\alpha,\beta} \gamma_{\alpha\beta}(\omega_t)\bigg(A_\beta(\omega_t)\mathcal{M}_t(.)A^\dagger_\alpha (\omega_t)-\frac{1}{2}\lbrace A^\dagger_\alpha (\omega_t)A_\beta(\omega_t),\mathcal{M}_t(.)\rbrace\bigg), \\
\nonumber&=\mathcal{M}_t \, 
\blue{\left[ \sum_{\omega_t} \sum_{\alpha,\beta} \gamma_{\alpha\beta}(\omega_t)e^{-\beta\omega_t}A_\beta(\omega_t)(.)A^\dagger_\alpha (\omega_t)\right]}
-\frac{1}{2}\mathcal{M}_t \,
\blue{\left[\sum_{\omega_t} \sum_{\alpha,\beta} \gamma_{\alpha\beta}(\omega_t)\lbrace A^\dagger_\alpha (\omega_t)A_\beta(\omega_t),(.)\rbrace \right]}, \\
\nonumber&=\mathcal{M}_t  \, 
\blue{\left[ \sum_{\omega_t} \sum_{\alpha,\beta} \gamma_{\beta \alpha}(-\omega_t)A^\dagger_\beta(-\omega_t)(.)A_\alpha (-\omega_t)\right]}
-\frac{1}{2}\mathcal{M}_t \, 
\blue{\left[\sum_{\omega_t} \sum_{\alpha,\beta} \gamma_{\alpha\beta}(\omega_t)\lbrace A^\dagger_\alpha (\omega_t)A_\beta(\omega_t),(.)\rbrace \right]}, \\
\nonumber&=\mathcal{M}_t \, 
\blue{\left[\sum_{\omega_t} \sum_{\alpha,\beta} \gamma_{\alpha \beta }(\omega_t)A^\dagger_\alpha(\omega_t)(.)A_\beta (\omega_t)\right]} 
-\frac{1}{2}\mathcal{M}_t \, 
\blue{\left[\sum_{\omega_t} \sum_{\alpha,\beta} \gamma_{\alpha\beta}(\omega_t)\lbrace A^\dagger_\alpha (\omega_t)A_\beta(\omega_t),(.)\rbrace\right]}, \\
&=\mathcal{M}_t \, 
\blue{\left[\mathcal{D}_t^\dagger(.)\right]} ,
\end{align}
where in the second line we used~\eqref{eq:DB3}, in the third line we used~\eqref{eq:Aw} and~\eqref{eq:KMS}, in the fourth line swapped indices $-\omega_t\to \omega_t$ and $\alpha\to\beta$, and in the final line used the definition of the adjoint superoperator $\Tr{\mathcal{D}_t^\dagger(B^\dagger) A}=\Tr{B^\dagger \mathcal{D}_t(A)}$ and the fact that the second term is self-adjoint. We next introduce a \blue{complementary} Lindbladian of form 
\begin{align}
\tilde{\mathscr{L}}_t(.)=\mathscr{U}_t(.)+\mathcal{D}_t^\dagger(.).
\end{align}
Notably the real part of the spectrum of $\tilde{\mathscr{L}}_t(.)$ coincides with that of the original Lindbladian $\mathcal{L}_t(.)$ due to the fact that $\mathcal{U}_t(.)$ is skew hermitian. \blue{Recalling that by assumption $\pi_t$ is a unique fixed point of $\mathcal{L}_t(.)$, this implies that $\tilde{\mathscr{L}}_t(.)$ also has a unique fixed point $\tilde{\pi}_t$.} We can thus define a pair of Drazin inverses given by
\begin{align}
&\mathcal{L}_t^+(A)=-\int^\infty_0 d\nu \ e^{\nu \mathcal{L}_t}(A-\Tr{A}\pi_t), \\
&\tilde{\mathcal{L}}_t^+(A)=-\int^\infty_0 d\nu \ e^{\nu \tilde{\mathcal{L}}_t}(A-\Tr{A}\tilde{\pi}_t).
\end{align}
for any $A\in M_d$. These inverses act according to
\begin{align}\label{eq:inverse1}
\nonumber&\mathscr{L}_t\mathscr{L}_t^+(A)=\mathscr{L}_t^+\mathscr{L}_t(A)=A-\pi_t\Tr{A}, \\
&\tilde{\mathscr{L}}_t\tilde{\mathscr{L}}_t^+(A)=\tilde{\mathscr{L}}_t^+\tilde{\mathscr{L}}_t(A)=A-\tilde{\pi}_t\Tr{A}.
\end{align}
By using~\eqref{eq:comm2} and~\eqref{eq:comm3} we have
\begin{align}\label{eq:inverse2}
\mathscr{L}_t\mathcal{M}_t(A)=\mathcal{M}_t\tilde{\mathscr{L}}_t(A).
\end{align}
For any traceless matrix $\lbrace B \ | \ B\in M_d, \ \Tr{B}=0 \rbrace$, we can combine~\eqref{eq:inverse1} and~\eqref{eq:inverse2} to get
\begin{align}\label{eq:inverse3}
\mathscr{L}_t^+\mathcal{M}_t(B)=\mathcal{M}_t\tilde{\mathscr{L}}_t^+(B),
\end{align}
We also define the following superoperator:
\begin{align}
\mathcal{V}_t(.):=-\frac{\mathscr{L}_t^++[\tilde{\mathscr{L}}_t^+]^\dagger}{2}(.).
\end{align}
Using~\eqref{eq:inverse3} one can also see that
\begin{align}\label{eq:lambda}
\mathcal{V}_t \mathcal{M}_t(B)=\mathcal{M}_t\mathcal{V}_t^\dagger (B).
\end{align}
Furthermore, since by assumption the real part of the eigenvalues of the Lindbladian $\mathscr{L}_t$ are negative, the same holds true for both Drazin inverses $\mathscr{L}_t^+$ and $\tilde{\mathscr{L}}_t^+$. \blue{To see this, let us consider any non-zero eigenvalue of $\mathscr{L}_t$ such as $z=x+i y$, with $x,y\in\Re e$. For the Drazin inverse $\mathscr{L}_t^+$, which shares the same eigenvectors as $\mathscr{L}_t$, the corresponding eigenvalue is given by $z^{-1}=(x+i y)^{-1}=(x-i y)/(x^2 + y^2)$ \cite{Boullion1971}. By assumption $x<0$, and thus all eigenvalues of $\mathscr{L}_t^+$ must also have a negative real part $\Re e(z^{-1})=x/(x^2 + y^2)<0$. We also note that the non-zero eigenvalues of $\tilde{\mathscr{L}}_t$ must also have a negative real part since the real part of the spectrum coincides with that of $\mathscr{L}_t$. By the same argument as above, this means that the non-zero eigenvalues of $\tilde{\mathscr{L}}^+_t$ have a negative real part. }

As a result, the eigenvalues of $\mathcal{V}_t(.)$ must have a positive real part. By Corollary 4.2 of \cite{Hassi2005}, a matrix product $XY$ with $Y\geq 0$ is positive if the eigenvalues of $X$ have no negative real part and $XY=YX^\dagger$. Since $\mathcal{M}_t(.)$ is positive,~\eqref{eq:lambda} implies that 
\begin{align}\label{eq:positive}
\mathcal{V}_t\mathcal{M}_t\geq 0.
\end{align}
Finally, we return to the trace functional~\eqref{eq:skew3}. Let us introduce the projection onto the traceless subspace $\mathcal{P}_T(A)=A-\Tr{A}\mathbb{I}/d$. Taking all results together one gets
\begin{align}\label{eq:positive2}
\nonumber\mathscr{I}(\pi_t,A)&=-\text{Re} \ \Tr{A \mathscr{L}_t^+ \mathcal{M}_t(A)}, \\
\nonumber&=-\text{Re} \ \Tr{\mathcal{P}_T(A) \mathscr{L}_t^+  \mathcal{M}_t\mathcal{P}_T(A)}, \\
\nonumber&=-\frac{1}{2} \Tr{B [\mathscr{L}_t^+ \mathcal{M}_t+\mathcal{M}_t[\mathscr{L}_t^+]^\dagger](B)}, \\
\nonumber&=-\frac{1}{2} \Tr{B [\mathscr{L}_t^+ +[\tilde{\mathscr{L}}_t^+]^\dagger]\mathcal{M}_t(B)}, \\
\nonumber&=\Tr{B \mathcal{V}_t \mathcal{M}_t(B)}, \\
&\geq 0,
\end{align}
where in the second line we used the fact that only traceless elements contribute to the functional due to~\eqref{eq:spectrum}, in the third line we set $\mathcal{P}_T(A)=B$ and $\mathcal{M}_t^\dagger(.)=\mathcal{M}_t(.)$, in the fourth line we used~\eqref{eq:inverse3} and in the final line we used the matrix positivity~\eqref{eq:positive}. Since the above holds for any hermitian matrix $A$, we conclude that the dynamical skew information is positive. Under time integration we therefore have $Q_w\geq 0$ and inequality Eq. (9) \blue{To conclude, we now prove $\mathscr{I}(\pi_t,A)=0$ if and only if $[A,\pi_t]=0$. Without loss of generality we may assume $\Tr{A}=0$. If $[A,\pi_t]=0$ then $\mathcal{M}_t(A)=0$ and thus $\mathscr{I}(\pi_t,A)=0$. On the other hand, let us instead suppose $[A,\pi_t]\neq 0$, in which case one necessarily has $\mathcal{M}_t(A)\neq 0$ which follows from~\eqref{eq:spectrum}. Now note that the unique fixed point of $\mathcal{V}_t$ is $\pi_t$. Clearly $\mathcal{M}_t(A)\not\propto \pi_t$ if $[A,\pi_t]\neq 0$, and hence $\mathcal{V}_t\mathcal{M}_t(A)\neq 0$. Using the positivity of $\mathcal{V}_t\mathcal{M}_t$ we then rewrite~\eqref{eq:positive2} as $\mathscr{I}(\pi_t,A)=\langle \sqrt{\mathcal{V}_t\mathcal{M}_t(A)},\sqrt{\mathcal{V}_t\mathcal{M}_t(A)} \rangle>0$. Therefore $\mathscr{I}(\pi_t,A)=0 \Leftrightarrow [A,\pi_t]=0$. This further implies that the total quantum correction $Q_w$ vanishes if and only if $[\pi_t,\dot{H}_t]=[H_t,\dot{H}_t]=0$ at all times. }

\section{Finding geodesics for a single parameter}\label{app:geodesic}

Consider a single parameter change $\lambda_0\to\lambda_\tau$ with Hamiltonian $H_t=X_0+\lambda_t X$, where in general $[X_0,X]\neq 0$. Furthermore, denote the rescaled work fluctuations by $\tilde{\sigma}^2_w=\frac{1}{2}\beta\sigma^2_w$. The aim is to minimise the linear objective function
\begin{align}
\mathcal{C}_\alpha:=\alpha\tilde{\sigma}_w^2+(1-\alpha)\Wdis; \ \ \ \alpha\in[0,1],
\end{align}
with respect to protocol $\lambda_t$. 
\blue{Using the metrics $\Lambda(\lambda_t)$ and $\xi(\lambda_t)$ from the main text in Eq. (13) and Eq. (14), we have 
\begin{align}\label{eq:object}
\mathcal{C}_\alpha
&=   \int_{0}^{\tau}dt\,\ \left(\alpha \dot{\lambda}_t \, {\Lambda}({\lambda}_t)   \, \dot{\lambda}_t +(1-\alpha) \, \dot{\lambda}_t \, {\xi}({\lambda}_t) \, \dot{\lambda}_t \right),
=\int^\tau_0 dt \ \dot{\lambda}_t^2\big(  \xi (\lambda_t)+\alpha\beta \diffX \big),
\end{align}
where $\beta \diffX = {\Lambda}({\lambda}_t)  -  {\xi}({\lambda}_t)$ is the dynamical skew information for a single parameter change (which depends on $\lambda_t$ through $\pi_t$), following from the general definition of $\diffA$ in the main text before Eq.~(10). 
The functional~\eqref{eq:object} is minimised by the solution to the Euler-Lagrange equation for the cost function $C_\alpha(\lambda_t,\dot{\lambda}_t):=\dot{\lambda}_t^2\big(  \xi (\lambda_t)+\alpha\beta\mathcal{I}_t(\pi_t,X)\big)$, i.e.
\begin{align}\label{eq:EL}
\frac{\partial C_\alpha}{\partial \lambda_t}=\frac{d}{dt}\bigg[\frac{\partial C_\alpha}{\partial \dot{\lambda}_t}\bigg],
\end{align}
which gives
\begin{align}
\dot{\lambda}_t^2 \frac{d }{d \lambda_t} \big(  \xi (\lambda_t)+\alpha\beta\mathcal{I}_t(\pi_t,X)\big)
&=\frac{d}{dt}\bigg[2  \dot{\lambda}_t \big(  \xi (\lambda_t)+\alpha\beta\mathcal{I}_t(\pi_t,X)\big) \bigg], \nonumber \\
&= 2  \ddot{\lambda}_t \big(  \xi (\lambda_t)+\alpha\beta\mathcal{I}_t(\pi_t,X)\big) + 2  \dot{\lambda}_t^2 \frac{d}{d\lambda_t}\bigg[ \big(  \xi (\lambda_t)+\alpha\beta\mathcal{I}_t(\pi_t,X)\big) \bigg]. \nonumber \\
\Longrightarrow 0 &= 2  \ddot{\lambda}_t \big(  \xi (\lambda_t)+\alpha\beta\mathcal{I}_t(\pi_t,X)\big) +   \dot{\lambda}_t^2 \frac{d}{d\lambda_t}\bigg[ \big(  \xi (\lambda_t)+\alpha\beta\mathcal{I}_t(\pi_t,X)\big) \bigg], \nonumber \\
\Longrightarrow \ddot{\lambda}_t&=-\frac{\partial C_\alpha(\lambda_t,\dot{\lambda}_t)}{\partial \lambda_t}\frac{1}{C_\alpha(\lambda_t,\dot{\lambda}_t)} {\dot{\lambda}_t^2 \over 2}.
\end{align}
}
 Solving~\eqref{eq:EL} yields an equation for the optimal velocity of the control parameter for a given $\alpha$:
\begin{align}\label{eq:optimalsol}
\dot{\lambda}^{\text{opt}}_t(\alpha)=\frac{(\lambda_\tau-\lambda_0)\big(\xi (\lambda_t)+\alpha\beta\mathcal{I}_t(\pi_t,X)\big)^{-1/2}}{\int^\tau_0  dt \ \big(\xi (\lambda_t)+\alpha\beta\mathcal{I}_t(\pi_t,X)\big)^{-1/2} },
\end{align}
One concludes that the optimal velocity is proportional to the following:
\begin{align}
\dot{\lambda}^{\text{opt}}_t(\alpha)\propto \big(\xi (\lambda_t)+\alpha\beta\mathcal{I}_t(\pi_t,X)\big)^{-1/2}.
\end{align}

\section{Thermodynamic metrics for the harmonic oscillator}\label{app:HarmonicOscillator}

We wish to evaluate the following metrics for the single parameter $\omega_t$:
\begin{align}\label{eq:metrics1}
&\dot{\omega}^2_t\Lambda(\omega_t):= \frac{\beta}{\Gamma}\Tr{\dot{H} \  \mathbb{S}_{\pi_t}(\dot{H})}, \\
&\dot{\omega}^2_t\xi(\omega_t):= \frac{\beta}{\Gamma}\Tr{\dot{H} \  \mathbb{J}_{\pi_t}(\dot{H})}.
\end{align}
For the harmonic oscillator the Hamiltonian and power operator are given by
\begin{align}
	\nonumber&H_t = \hbar \omega_t (n_t + {1\over 2}), \\
	&\dot{H}_t= \left(\frac{\dot{\omega}_t}{\omega_t}\right) \bigg(H_t+\hbar\omega_t\frac{(a^\dagger_{\omega_t})^2+a^2_{\omega_t}}{2}\bigg),
\end{align}
where $n_t = a_{\omega_t}^\dag a_{\omega_t}$ with $a_{\omega_t} = \sqrt{m \omega_t \over 2 \hbar}(x + i {p \over m \omega_t})$. The metrics~\eqref{eq:metrics1} can then be simplified as 
\begin{align}\label{eq:metrics2}
&\Lambda(\omega_t)= \frac{\beta}{\Gamma} \left( {1 \over \omega_t} \right)^2 \, \left( \Tr{H^2_t \pi_t}  + \Tr{(\delta H_t)^2  \pi_t}  -  \left(\Tr{H_t \pi_t}\right)^2 \right), \\
&\xi(\omega_t)= \frac{\beta}{\Gamma}\left({1\over \omega_t}\right)^2 \left( \int_0^1 \d a \, \Tr{\delta H_t  \, \pi_t^a \, \delta H_t  \, \pi_t^{1-a}}+ \Tr{H_t^2  \, \pi_t}  -  \left(\Tr{H_t \, \pi_t}  \right)^2 \right).
\end{align}
where we introduce $\delta H_t=\hbar\omega_t((a^\dagger_{\omega_t})^2+a^2_{\omega_t})/2$ for shorthand. By working in the number basis $H_t=\hbar\omega_t\sum_{n=0}^\infty( n_t+\frac{1}{2})\ket{n}\bra{n}$ and using the standard relations $a_{\omega_t}^\dagger \ket{n_t}=\sqrt{n_t+1}\ket{n_t+1}$ and $a_{\omega_t}\ket{n_t}=\sqrt{n_t}\ket{n_t-1}$, a textbook calculation reveals the following expressions:
\begin{align}
&\left(\Tr{H_t \, \pi_t} \right)^2 =  (\hbar \omega_t)^2  \, {1 + 2 e^x + e^{2x} \over 4 (e^x - 1)^2}, \\
&\Tr{H_t^2 \, \pi_t}=  (\hbar \omega_t)^2 \, {1 + 6 e^x + e^{2x} \over 4 (e^x - 1)^2},\\
&\Tr{(\delta H_t)^2  \pi_t}= \left( \hbar \omega_t \right)^2   \, {2 e^{2x} + 2 \over 4 (e^x -1)^2}, \\
&\Tr{\delta H_t  \, \pi_t^a \, \delta H_t  \, \pi_t^{1-a}} ={(\hbar \omega_t)^2 \over 4} \,  {2e^{2x} \over  (e^x - 1)^2}  \left(e^{- 2x (1-a)}  + e^{- 2x a }  \right),
\end{align} 
where $x=\beta\hbar\omega_t$. After plugging these expressions into~\eqref{eq:metrics2} we get
\begin{align}\label{eq:metricsosc1}
	\Lambda(\omega_t) &=   { \beta \hbar^2   \over  4 \Gamma \sinh^2 \left({\beta \hbar \omega_t / 2}\right)}  \, \left( 1 + \cosh \beta \hbar \omega_t   \right) , \\
	\xi(\omega_t) &=  {\beta \hbar^2      \over  4 \Gamma \sinh^2 \left({\beta \hbar \omega_t / 2}\right)} \, \left(1 +  {\sinh \beta \hbar \omega_t  \over \beta \hbar \omega_t}\right). \label{eq:metricsosc2}
\end{align}
These quantities are used to plot Fig. 1(a)-(b) in the main text. 

\blue{From this we can analyse the limiting behaviour of the metrics in different temperature regimes. In the limit $\beta \to 0$ these metrics reduce to 
\begin{align}
	\Lambda(\omega_t)  \approx   {1   \over  \Gamma  \omega_t^2}   \, \left( {2 \over \beta} + {\beta  \over 2}  \, (\hbar \omega_t )^2 \right) 
	\quad	\mbox{ and } \quad
	\xi(\omega_t) \approx   {1   \over  \Gamma  \omega_t^2}   \, \left( {2 \over \beta} + {\beta  \over 3}  \, (\hbar \omega_t )^2 \right) ,
\end{align}
so that
\begin{align}
	\lim_{\beta \to 0} \left( \Lambda(\omega_t) - \xi(\omega_t) \right)  =  {\hbar^2  \over  \Gamma }    	\lim_{\beta \to 0}  \left( {\beta  \over 2} - {\beta  \over 3}  \right) =0.
\end{align}
In the limit $\beta \to \infty$ the above metrics reduce to 
\begin{align}
	\Lambda(\omega_t) 
	%\approx   { \beta \hbar^2 \, \exp \left(- {\beta \hbar \omega_t }\right)   \over   \Gamma}  \, \left( 1 + {\exp(\beta \hbar \omega_t)  \over 2 } \right) 
	\approx  \beta \,  { \hbar^2  \over  2 \Gamma} 
	\quad	\mbox{ and } \quad
	\xi(\omega_t) 
	%&\approx { \beta \hbar^2  \, \exp \left(- {\beta \hbar \omega_t }\right)  \over   \Gamma }  \, \left(1 +  {\exp(\beta \hbar \omega_t)  \over 2 \beta \hbar \omega_t}\right) 
		\approx   { \hbar  \over  2 \Gamma   \omega_t} .
\end{align}
Therefore in the high temperature limit ($\beta\rightarrow 0$), the metrics $\xi(\omega_t)$ and $\Lambda(\omega_t)$ become equal, indicating vanishing of the quantum fluctuations $Q_w$. On the other hand, when the temperature is low  ($\beta\rightarrow\infty$), the dissipation metric $\xi(\omega_t)$ converges to a constant in $\beta$, $\lim_{\beta\to\infty}\xi(\omega_t)=\hbar/2 \Gamma \omega_t$, while the fluctuation metric $\Lambda(\omega_t)$ grows linearly $\Lambda(\omega_t) \approx \beta \, \hbar^2/2 \Gamma$ for $\beta \to \infty$, see Fig. 1(b) in the main text. 
}

\medskip

Given a fixed initial and final frequency $(\omega_0, \omega_\tau)$, we can now use~\eqref{eq:metricsosc1} and~\eqref{eq:metricsosc2} to minimise the objective function
\begin{align}\label{eq:alphaObjective2}
	\nonumber\mathcal{C}_\alpha&=\alpha \, \tilde{\sigma}_w^{2}+(1-\alpha) \, \Wdis, \\
	&=\int^\tau_0 dt \ \dot{\omega}_t^2\big( \alpha \Lambda(\omega_t) +(1-\alpha)\xi (\omega_t)\big),
\end{align}
for any $\alpha\in[0,1]$. Let $\omega^{opt}_t=\omega^{opt}_t(\alpha)$ denote the optimal solution minimising~\eqref{eq:alphaObjective2} for a given $\alpha$. Using the solution~\eqref{eq:optimalsol}, we find an implicit equation for $\omega^{opt}_t(\alpha)$:
\begin{align}\label{eq:optimalsol2}
	{\dot{\omega_t} \over \omega_{\tau} - \omega_0} 
	=  {{ \left| \sinh {x\over 2} \right|  \over  \sqrt{1 + \alpha  \,  \cosh x  + (1-\alpha)   {\sinh x  \over x}} } \over \bigintss_0^{\tau} \d t  \,  { \left| \sinh {x\over 2} \right|  \over  \sqrt{1 + \alpha  \,  \cosh x  + (1-\alpha)   {\sinh x  \over x}} }  } .  
\end{align}
The numerical solutions for~\eqref{eq:optimalsol2} at each value of $\alpha$ are finally used to compute the points on the Pareto front presented in \newline Fig. 2 from the main text.

\section{Work fluctuation-dissipation relation for discrete processes beyond weak-coupling}

\blue{Our analysis in the main text assumes that at all times the system undergoes dissipative Markovian evolution, which relies on the assumption that the coupling to the bath is sufficiently weak. In this Appendix we present an alternative picture in which the system is subject to a series a fast changes in its Hamiltonian, with each quench proceeded by thermalisation with respect to the bath. However, here no restriction will be placed on the strength of coupling, and we will prove a quantum work FDR analogous to our main result Eq. (8). }

\blue{Let $\lbrace \sys{H}^{(1)}, \sys{H}^{(2)}, ...\sys{H}^{(N)} \rbrace$ represent a particular sequence of $N$ quenches in the system Hamiltonian, where at each step we take the total Hamiltonian to be of form
\begin{align}
\SB{H}^{(i)}:=\sys{H}^{(i)}\otimes \bath{\mathbb{I}}+\sys{\mathbb{I}}\otimes\bath{H}+\gamma\SB{V},
\end{align}
with $\SB{V}$ a time independent interaction with arbitrary coupling strength $\gamma$. The spectral decomposition of the total Hamiltonian is denoted by $\SB{H}^{(i)}=\sum_{n}\epsilon_n^{(i)} \ket{\epsilon_n^{(i)}}\bra{\epsilon_n^{(i)}}$. In addition to being initially thermal, at the end of the $(i-1)$'th quench we assume that: 
\begin{itemize}
\item the system and bath equilibrates to the time-averaged state, i.e., it decoheres  (see the review \cite{Gogolin_2016} for details)
\item the system relaxes to the reduced of a global thermal state, $\sys{\tilde{\pi}}^{(i)}=\tr_B(\SB{\pi}^{(i)})=\tr_B(e^{-\beta \SB{H}^{(i)}}/\SB{Z}^{(i)})$. Note that $\sys{\tilde{\pi}}^{(i)}$ deviates from a local Gibbs state unless the weak-coupling assumption is taken, namely if $\gamma^2\ll 1$. 
\end{itemize}
It is important to realise that we make weak assumptions on the state of B, and that we only assume that thermalisation takes place at the level of S (i.e. one can imagine that SB evolve unitarily leading to thermalisation at the local level). Alternatively, one may assume that SB is put in contact via weak coupling to an external super-bath at inverse temperature $\beta$, leading to thermalisation of the full SB, but this stronger requirement is not needed for our derivation. }

\blue{Crucially,  each change $\sys{H}^{(i)}\to \sys{H}^{(i+1)}$ is sufficiently fast such that the system-bath state does not change, while the thermalisation stages occur with no work done on the system due to the fixed Hamiltonian. Since work is performed only during the quench stages, with evolution effectively unitary, the resulting work distribution can be obtained by successive iterations of the two-projective measurement protocol applied to system and bath at each stage. By treating $w=\sum^{N-1}_{i=1}w^{(i)}$ as the total work done along the N steps, which is composed of a sum of independent random variables, the work distribution is given by
\begin{align}
P(w):=\prod^{N-1}_{i=1} P(w^{(i)}); \ \ \ \ \ P(w^{(i)}):=\sum_{n,m} \delta [w^{(i)}-\epsilon^{(i+1)}+\epsilon^{(i)}] \big| \braket{\epsilon_m^{(i+1)}| \epsilon_n^{(i)}} \big|^2 \bra{\epsilon_n^{(i)}} \SB{\pi}^{(i)} \ket{\epsilon_n^{(i)}}. 
\end{align} 
The first and second cumulants of work are subsequently given by 
\begin{align}
&\langle w \rangle =\sum^{N-1}_{i=1} \TrS{(\sys{H}^{(i+1)}-\sys{H}^{(i)})\sys{\tilde{\pi}}^{(i)}}, \\
\label{eq:workfluct1}&\sigma_w^2=\sum^{N-1}_{i=1} \TrS{(\sys{H}^{(i+1)}-\sys{H}^{(i)})^2\sys{\tilde{\pi}}^{(i)}}-\TrS{(\sys{H}^{(i+1)}-\sys{H}^{(i)})\sys{\tilde{\pi}}^{(i)}}^2.
\end{align}
Note that the above terms depend only on the system degrees of freedom since each quench applies only locally to the system Hamiltonian. }

\blue{Our focus will be on slow processes, which in this context implies that the number of steps $N$ is sufficiently large. By re-expressing $\langle w \rangle$  in terms of SB as: $\langle w \rangle =\sum^{N-1}_{i=1} \Tr{(\SB{H}^{(i+1)}-\SB{H}^{(i)})\pi_{SB}^{(i)}}$, it is straightforward to show that the dissipated work $\Wdis=\langle w \rangle -\Delta F$ can be rewritten as \cite{Gallego_2014,Perarnau_2018c}
\begin{align}\label{eq:wdiss1}
\Wdis=\frac{1}{\beta}\sum^{N-1}_{i=1} S(\SB{\pi}^{(i)} || \SB{\pi}^{(i+1)}),
\end{align}
where $S(\rho|| \sigma)=\Tr{\rho(\ln \rho -\ln \sigma)}$ is the quantum relative entropy. Note that for a density operator $\rho(t)$ that depends smoothly on some parameter $t$, the relative entropy between close states $\rho(t+\delta t)$ and $\rho(t)$ is approximated up to second order in $\delta t$ as follows \cite{Hayashi2002}:
\begin{align}\label{eq:relent}
S\big(\rho(t) ||\rho(t+\delta t)\big)=\frac{1}{2 }\delta t^2 \ \Tr{ \frac{\partial \ln \rho(t)}{\partial_t} \ \mathbb{J}_{\rho(t)}\bigg(\frac{\partial \ln \rho(t)}{\partial_t}\bigg)}+\mathcal{O}(\delta t^3),
\end{align}
where $\mathbb{J}_{\rho}(A)$ is defined in Eq. (9). By defining $\Delta \sys{H}^{(i)}/N=(\sys{H}^{(i+1)}-\sys{H}^{(i)})$ and identifying $1/N$ as a small parameter, the expansion~\eqref{eq:relent} yields the following approximation for $S(\SB{\pi}^{(i)} || \SB{\pi}^{(i+1)})$ after taking the partial trace over the bath degrees of freedom:
\begin{align}\label{eq:relent2}
S(\SB{\pi}^{(i)} || \SB{\pi}^{(i+1)})=\frac{\beta^2}{2 N^2} \TrS{\Delta \sys{H}^{(i)} \ \mathbb{J}_{\sys{\tilde{\pi}}^{(i)}}(\Delta \sys{H}^{(i)})}+\mathcal{O}(1/N^3).
\end{align}
At this stage we introduce a one-parameter family of Hamiltonians $\lbrace \sys{H}(t) \ | \ t\in[0,1], \ \sys{H}(t=i/N)=\sys{H}^{(i)} \rbrace$ and denote the corresponding system state by 
\begin{align}
\sys{\tilde{\pi}}(t)=\frac{\TrB{e^{-\beta (\sys{H}(t)+\bath{H}+\gamma\SB{V})}}}{\TrSB{e^{-\beta (\sys{H}(t)+\bath{H}+\gamma\SB{V})}}},
\end{align}
Then one has $\Delta \sys{H}^{(i)}=\sys{\dot{H}}(i/N)+\mathcal{O}(1/N)$ where $\sys{\dot{H}}(t)=\frac{\partial }{\partial t} \sys{H}(t)$ represents the system's power operator. Combining~\eqref{eq:wdiss1} and~\eqref{eq:relent2} and taking the continuum limit gives 
\begin{align}\label{eq:wdiss2}
\Wdis=\frac{\beta}{2N}\int^1_0 dt \ \TrS{ \sys{\dot{H}}(t) \ \mathbb{J}_{\sys{\tilde{\pi}}(t)}(\sys{\dot{H}}(t))}+\mathcal{O}(1/N^2).
\end{align}
This gives the dissipated work in the limit of many discrete steps up to first order. }

\blue{We now turn to evaluating the work fluctuations $\sigma_w^2=\langle w^2 \rangle-\langle w \rangle^2$ in this limit. From~\eqref{eq:workfluct1} we find in the continuum limit
\begin{align}\label{eq:workfluct2}
\nonumber \sigma_w^2&=\frac{1}{N^2}\sum^{N-1}_{i=1} \TrS{(\Delta \sys{H}^{(i)})^2\sys{\tilde{\pi}}^{(i)}}-\frac{1}{N^2}\sum^{N-1}_{i=1}\TrS{\Delta \sys{H}^{(i)}\sys{\tilde{\pi}}^{(i)}}^2, \\
&=\frac{1}{N}\int^1_0 dt \ \bigg(\Tr{\sys{\dot{H}}^2(t) \ \sys{\tilde{\pi}}(t)}-\TrS{\sys{\dot{H}}(t) \ \sys{\tilde{\pi}}(t)}^2\bigg)+\mathcal{O}(1/N^2), 
\end{align}
We are now ready to obtain the work fluctuation-dissipation relation for discrete processes. For a large number of steps $N^2\gg 1$, subtracting~\eqref{eq:wdiss2} from~\eqref{eq:workfluct2} gives
\begin{align}\label{eq:qFDR2}
\Wdis=\frac{1}{2}\beta \sigma_w^2- Q_w,
\end{align}
where 
\begin{align}
Q_w=\frac{\beta}{2N}\int^1_0 dt \ \mathcal{I}\big(\sys{\tilde{\pi}}(t),\sys{\dot{H}}(t)\big),
\end{align}
is the time-integrated average Wigner-Yanase-Dyson skew information, defined as in the main text by
\begin{align}
\mathcal{I}(\rho,A):=-\frac{1}{2}\int^1_0 da \ \Tr{[A,\rho^a][A,\rho^{1-a}]}.
\end{align}
It follows from the positivity of the skew information that
\begin{align}
\Wdis\leq \frac{1}{2}\beta \sigma_w^2,
\end{align}
with equality \textit{iff} $[\sys{\dot{H}}(t),\sys{\tilde{\pi}}(t)]=0 \ \ \forall t\in[0,1]$. }

\blue{Comparing~\eqref{eq:qFDR2} with Eq. (8) obtained in the Lindblad approach, we see structural similarities. Here one finds a quantum correction term stemming from the fact that the reduced state of the system does not necessarily commute with the power operator $\dot{\sys{H}}(t)$. In the weak coupling limit where $\tilde{\sys{\pi}}\simeq \sys{\pi}$ we recover the same quantum correction term for Lindblad equation Eq. (10), with $N=\tau \Gamma/2$ playing the role of the ratio between the evolution and equilibration timescales. Clearly we see that the resulting work distribution has an increased spread due to additional quantum fluctuations in power, quantified by the skew information. This suggests that the assumption of weak coupling used to obtain Eq. (8) in the main text is not crucial, and that the result is more fundamentally linked to the linear expansion of the system state close to the quasi-static limit. }

\section{Numerical verification of Eq. (8)}

\blue{In order to show how the slow driving approximation affects the FDR we numerically simulate here the exact dynamics of a two level system in contact with a bosonic thermal bath with flat spectrum. The Lindblad equation we use in the eigenbasis of the system Hamiltonian is given by~\cite{breuer2002theory}:
\begin{align}
\dot\rho_t &= \lind_r\sqrbra{ \rho} =\gamma (P_r+1)\left(\hat\sigma_-\rho_t\hat\sigma_+ - \frac{1}{2}\curbra{\hat\sigma_+\hat\sigma_-,\rho_t}\right) +\gamma P_r\left(\hat\sigma_+\rho_t\hat\sigma_- - \frac{1}{2}\curbra{\hat\sigma_-\hat\sigma_+,\rho_t}\right),
\label{eq:lindbladian}
\end{align}
where $r$ is the time dependent energy spacing and $P_r$ is simply the Planck distribution $P_r := \frac{1}{e^{2 \beta r}-1}$. For simplicity, we assume the proper equilibration timescale $\gamma$ to be one. Since we have to account for the explicit change of Hamiltonian we define the unitary transformation on $SU(2)$: 
\begin{align}
	U_t := \{U_t \in SU(2) |\; U_t^\dagger H_t U_t \;\text{is diagonal}\}.
\end{align}
Notice that $SU(2)$ can be parametrised by spherical angles; for this reason we write the qubit Hamiltonian in spherical coordinates as:
\begin{align}
H_t = r_t\cos{\varphi_t}\sin{\theta_t} \,\hat \sigma_x+  r_t\sin{\varphi_t}\sin{\theta_t} \, \hat\sigma_y 
+  r_t\cos{ \theta_t} \, \hat\sigma_z = U(\theta_t, \phi_t) \;H_{r_t} \;U(\theta_t, \phi_t)^\dagger.
\end{align}
Then in this diagonal representation the Lindbladian equation is given as:
\begin{align}
	\dot{\rho}_t =  U(\theta_t, \phi_t)\lind\sqrbra{ U(\theta_t, \phi_t)^\dagger \rho _t U(\theta_t, \phi_t)}U(\theta_t, \phi_t)^\dagger.
\end{align}
The formal solution is given by 
\begin{align}\label{eq:solution}
\rho_t=\overleftarrow{\mathcal{T}} e^{\int^\tau_0 dt \ \mathscr{L}_t}(\rho_0); \ \ \ \ \ \ \ \ \  \rho_0=\frac{e^{-\beta H_0}}{Z_0}.
\end{align}
After determining the solution~\eqref{eq:solution} numerically, we can compute the first two moments of work using the following exact expressions from Appendix A:
\begin{align}
\label{eq:exactwdiss}&\Wdis=\beta^{-1}\ln \bigg(\frac{Z_\tau}{Z_0}\bigg)+\int^\tau_0 dt \ \Tr{\dot{H}_t \rho_t}, \\
\label{eq:exactfluc}&\sigma_w^2= 2\int^\tau_0\hspace{-2mm} dt_{1} \int^{t_{1}}_0 \hspace{-2mm}dt_{2} \ \Tr{\dot{H}_{t_{1}}\overleftarrow{P}(t_{1},t_{2})\left[\operatorS_{\varrho_{t_{2}}}(\dot{H}_{t_{2}})\right]}.
\end{align}
These quantities are shown in Figure 3 for two different protocols as a function of the total duration $\tau$. We clearly see that our slow driving approximations (ie. $1/\tau^2\ll 1$) becomes valid at large values of $\tau$, thereby verifying our work FDR Eq. (8). In the first case we consider a commuting protocol where only the energy levels are changed in time, which in spherical coordinates is simply given by 
\begin{align}\label{eq:classicalproto}
\lbrace r,\phi,\theta,\phi \rbrace (t)=\lbrace t,0,0 \rbrace.
\end{align}
The Hamiltonian stays in the same energy basis throughout the protocol, and thus the quantum correction $Q_w=0$. In the large $\tau$ limit we see that work dissipation and fluctuations converge to the classical FDR $\Wdis=\frac{1}{2}\beta \sigma_w^2$ as expected.}

\blue{For the second case we rotate the Hamiltonian energy basis using the following protocol:
\begin{align}\label{eq:quantumproto}
\lbrace r,\phi,\theta \rbrace (t)=\big\lbrace \sqrt{t^2-2t+2}, \ 0, \ \arctan(1/(1-t)) \big\rbrace.
\end{align}
As a result we generate coherence during the process and hence obtain a non-zero quantum correction $Q_w$. In Figure 3 one sees that again the slow driving approximations for $\Wdis$ and $\frac{1}{2}\beta\sigma_w^2$ are valid in the large $\tau$ limit. In contrast to the commuting protocol, at large $\tau$ the work fluctuations exceed the dissipation, confirming the inequality Eq. (9) in the main text. The difference between the curves for  $\frac{1}{2}\beta\sigma_w^2$ and $\Wdis$ represents the time-integrated skew information $Q_w$. Interestingly, we see that $\Wdis\leq\frac{1}{2}\beta \sigma_w^2$ remains valid at all times. It is not clear whether or not this is a general feature of the Lindblad master equation, or if this is simply a property of this particular spin-boson model.    }

\begin{figure}
	\centering
	\includegraphics[width=0.4\linewidth]{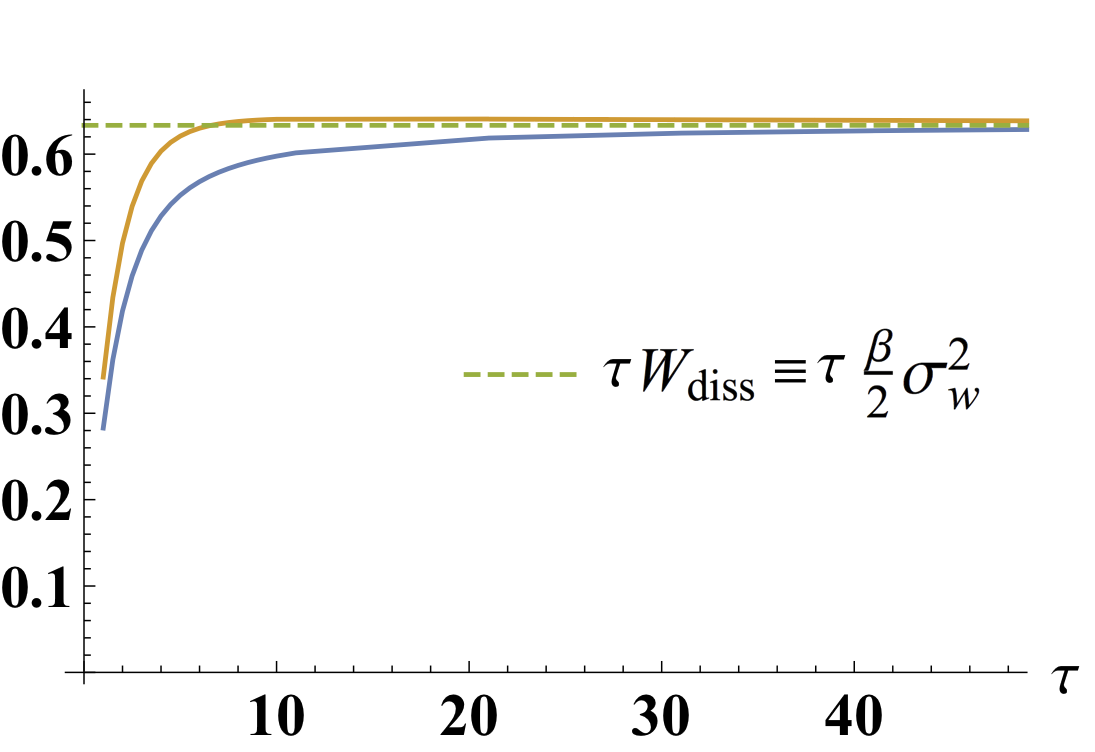}
	\quad\includegraphics[width=0.4\linewidth]{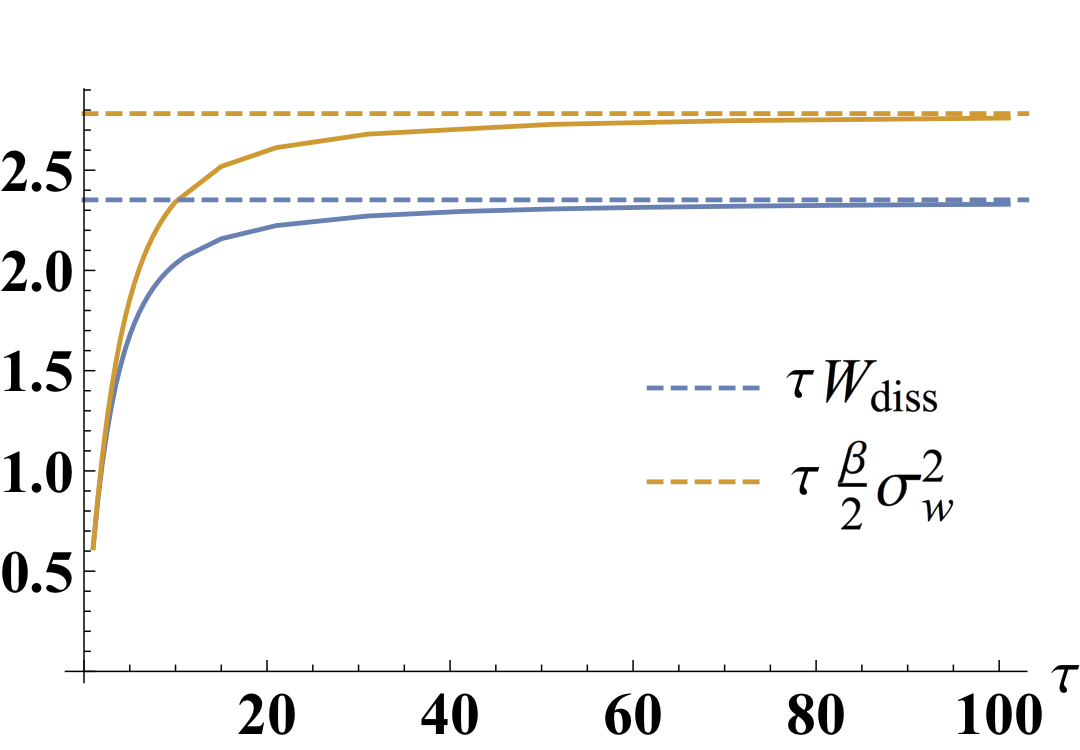}
	%\captionsetup{labelformat=empty}
	\caption{ The behaviour of average dissipation and work fluctuations for the two protocols: (Left) a \textit{classical} protocol where only the energy spacing is changed according to~\eqref{eq:classicalproto}, and (Right) a \textit{quantum} protocol that generates coherences via~\eqref{eq:quantumproto}. In both plots the solid lines represent numerical computations of the exact quantities~\eqref{eq:exactwdiss} and~\eqref{eq:exactfluc}, while the dashed lines are the theoretical results Eq. (4) and Eq. (6), computed using the slow driving approximation. }
\end{figure}

\end{document}